\documentclass[onecolumn]{autart}

\usepackage{lineno,cite}
\usepackage{amsmath,amssymb,amsfonts,mathrsfs}
\usepackage{epsfig,epstopdf}
\usepackage{graphicx,graphics}
\usepackage{color}
\usepackage{float}
\usepackage{ifpdf}
\usepackage[colorlinks=true]{hyperref}

\def\mbb{\mathbb}
\newtheorem{theorem}{Theorem}

\newtheorem{example}{Example}
\newtheorem{lemma}{Lemma}

\newtheorem{remark}{Remark}
\input{tcilatex}
\begin{document}


\begin{frontmatter}

\title{Dynamical analysis of  quantum linear systems driven by multi-channel multi-photon states\thanksref{footnoteinfo}}

\thanks[footnoteinfo]{This paper was not presented at any IFAC 
meeting. Corresponding author Guofeng Zhang. Tel. +852-2766-6936. 
Fax +852-2764-4382. Email: Guofeng.Zhang@polyu.edu.hk. This research is supported in part by Hong Kong RGC grants (Nos. 531213 and 15206915) and a National Natural Science Foundation of China (NSFC) grant (No. 61374057).}

\author{Guofeng Zhang}

\address{Department of Applied Mathematics, The Hong Kong Polytechnic University, Hung Hom, Kowloon,  Hong Kong}


\begin{abstract}
In this paper, we investigate the dynamics of quantum linear systems where the input signals are multi-channel multi-photon states, namely states determined by a definite number of photons superposed in multiple input channels. In contrast to most existing studies on separable input states in the literature, we allow the existence of quantum correlation (for example quantum entanglement) in these multi-channel multi-photon input states. Due to the prevalence of quantum correlations in the quantum regime, the results presented in this paper are very general. Moreover, the multi-channel multi-photon states studied here are reasonably mathematically tractable. Three types of multi-photon states are considered: 1) $m$ photons superposed among $m$ channels, 2) $N$ photons superposed among $m$ channels where $N\geq m$, and 3) $N$ photons superposed among $m$ channels where $N$ is an arbitrary positive integer.  Formulae for  intensities and states of output fields are derived. Examples are used to demonstrate the effectiveness of the results.  
\end{abstract}
\end{frontmatter}

\section{Introduction}\label{sec:introduction}

Dynamical response analysis is an essential ingredient of control
engineering, and is also the basis of controller design. For example,
impulse response, step response, and frequency response are standard
materials in modern control textbooks, see, e.g., \cite{KS72}, \cite{AM79}, 
\cite{ZDG96}, \cite{QZ09}. Fluctuation analysis of a dynamical
system driven by white noise underlies the celebrated Kalman filter and
linear quadratic Gaussian (LQG) control. Likewise, in the quantum regime,
the response of quantum linear systems to quantum Gaussian white noise is
the basis of quantum filtering and measurement-based feedback control, see,
e.g., \cite{VPB80}, \cite{VPB83}, \cite{HP84}, \cite{VPB92}, \cite{VPB93}, 
\cite{vHSM05}, \cite{WM10}, \cite{DP10}, \cite{AT12}, \cite{ZWL12}, \cite{ZLW+15}, \cite{GY16} and references therein.

In addition to quantum Gaussian noise commonly dealt with in quantum optical
laboratories, in recent years, highly nonclassical signals such as
single-photon states, multi-photon states, and Schr\"{o}dinger's cat states
have been attracting growing interest due to their promising applications
in quantum information technology \cite{NC00}. Roughly speaking, an $\ell $-photon state of a light beam means that the light field contains \textit{exactly} $\ell $ photons. In this paper, we are concerned with
continuous-mode $\ell $-photon states, that is, these photons are specified
by their frequency profiles centered at the carrier frequency of the light
field. Continuous-mode single- and multi-photon states have found important
applications in quantum computing, quantum communication, and quantum
metrology, see, e.g., \cite{GEP+98}, \cite{NC00}, \cite{RL00}, \cite{Ou07}, \cite{Milburn08}, \cite{CMR09}, \cite{GJN11}, \cite{GJNC12}, \cite{BCB+12}, \cite{BDS+12}, \cite{Yukawa:13}, \cite{zhang13}, \cite{zhang14}, \cite{YJ14}, 
\cite{NKM+15}.

In the quantum control community, the response of quantum systems to single-
and multi-photon states has been studied in the past few years. The
phenomenon of cross phase shift on a coherent signal induced by a single
photon pulse was investigated in \cite{MNM10}. Gough \textit{et al.} derived
quantum filters for Markov quantum systems driven by single-photon states or
Schr\"{o}dinger's cat states \cite{GJN11}, \cite{GJNC12}. The theory in \cite{GJN11} and \cite{GJNC12} was applied to the study of phase modulation in 
\cite{CHJ12}. Quantum master equations for an arbitrary quantum system
driven by multi-photon states were derived in \cite{BCB+12}. Quantum filters
for multi-photon states were derived in \cite{SZX15}, for both homodyne
detection and photodetection measurements. Numerical simulations carried out
in \cite{SZX15} for a two-level system driven by a 2-photon state revealed
interesting and complicated nonlinear behavior in the photon-atom
interaction. When a two-level atom, initialized in the ground state, is
driven by a single photon, the exact form of the output field state was
derived in \cite{PZJ15}. More discussions can be found in, e.g., \cite{FKS10}, \cite{LL13}, \cite{NKM+15} and references therein.

In \cite{zhang13}, an analytic expression of the output field state of a
quantum linear system driven by a single-photon state was derived.  The research initialized in \cite{zhang13} was continued and extended in \cite{zhang14}, where multi-photon states were considered. Unfortunately, the multi-photon states studied in  \cite{zhang14} are either with very limited quantum correlation or mathematically formidable. More specifically, the multi-photon states defined in \cite[Eqs. (23) and (41)]{zhang14} are separable states, i.e., there exists no entanglement among the channels.  A class of photon-Gaussian states was defined in \cite[Eq. (34)]{zhang14}. On the one hand, this class of states appears mathematically complicated. On the other hand, because each pulse shape is indexed by three parameters only,  the feature of the multi-channel entanglement is unclear.  A class of multi-channel multi-photon states was defined in \cite[Eq. (43)]{zhang14}, which, due to the presence of an $m$-fold product,  looks rather complicated mathematically.

The purpose of this paper is to provide a direct study of the dynamical
response of quantum linear systems to initially entangled multi-channel
multi-photon states. Unlike those separable  states studied in \cite{zhang13} and \cite{zhang14}, the multi-channel multi-photon states proposed in this paper are able to capture the entanglement among channels.  Examples presented in this paper demonstrate that these types
of multi-channel multi-photon states can be easily processed by quantum linear
systems. Furthermore, the proposed multi-channel multi-photon states are very general
as they contain many types of multi-channel multi-photon states as special
cases, see, e.g., \cite[Chapter 6]{RL00}, \cite{RMS07}, \cite{BDS+15}.
Finally, these states are mathematically more tractable than those in \cite[%
Eqs. (34) and (43)]{zhang14}. Therefore, the study carried out in this paper is more
relevant to quantum linear feedback networks and control.

Three types of multi-channel multi-photon states are studied in this paper.
Case 1): $m$ photons are superposed among $m$ channels. Specifically, the $m$-channel $m$-photon states are defined in Subsection \ref{subsec:m_photon}.
When the underlying quantum linear system is passive, an analytic
expression of the output intensity is presented in Subsection \ref{subsec:intensity}, see Theorem \ref{thm:intensity}. Moreover, the
steady-state output field state is investigated in Subsections \ref{sec:passive_mm} and \ref{subsec:invariant}, see Theorems \ref{thm:passive_mm} and \ref{thm:passive_invariance}. When the underlying
quantum linear system is non-passive, the steady-state output field state is no
longer an $m$-channel $m$-photon state, an explicit form of the output field
state is given in Subsection \ref{subsec:nonpassive}, see Theorem \ref{thm:nonpassive}. Case 2): $N$ photons are superposed among $m$ channels
where $N\geq m$. For this case, we assume the underlying quantum linear
system is passive. The analytic expressions of the output field state are
derived, see Theorems \ref{thm:aug23} and \ref{thm:passive_invariance_general}. Case 3): $N$ photons are superposed among $m$ channels, where $N$ is an arbitrary positive integer.  Specifically, a class of $m$-channel $N$-photon states are first presented in Subsection \ref{subsec:mN_input}, then in
Subsection \ref{subsec:mN_output}, the steady-state output field state of a
quantum linear passive system driven by an $m$-channel $N$-photon input
state is derived, see Theorem \ref{thm:general}.

\emph{Notation}. The imaginary unit $\sqrt{-1}$ is denoted by $\mathrm{i}$.
Given a column vector of complex numbers or operators $x=[x_{1}~\cdots
~x_{k}]^{T}$, define a column vector $x^{\#} \triangleq [x_{1}^{\ast }~\cdots
~x_{k}^{\ast }]^{T}$, where the superscript ``$\ast $'' stands for complex
conjugation of a complex number or Hilbert space adjoint of an operator.
Define a row vector $x^{\dag }\triangleq (x^{\#})^{T}= [x_{1}^{\ast }~\cdots
~x_{k}^{\ast }]$. Define a 
doubled-up column vector $\breve{x}\triangleq \lbrack x^{T}~x^{\dag }]^{T}$.
Let $I_{k}$ be an identity matrix and $0_{k}$ a zero square matrix, both of
dimension $k$. Denote $J_{k}=\mathrm{diag}(I_{k},-I_{k})$.  Given a matrix $X\in \mathbb{C}^{2j\times 2k}$, define $X^{\flat
}\triangleq J_{k}X^{\dag }J_{j}$. Given a matrix $A$, let $A^{jk}$ denote
the entry on the $j$th row and $k$th column. Let $m$ be the number of input
channels. Let $n$ be the number of degrees of freedom of a given quantum
linear system, namely the number of quantum harmonic oscillators. The ket $|\phi \rangle $ denotes the initial state of the system of interest, and $|0\rangle $
stands for the vacuum state of free fields. The convolution of two functions 
$f$ and $g$ is denoted as $f\circledast g$. Given two matrices $U$, $V\in 
\mathbb{C}^{r\times k}$, define a doubled-up matrix $\Delta (U,V)\triangleq
\lbrack U~V;V^{\#}~U^{\#}]$. Given two operators $\mathcal{A}$ and $\mathcal{B}$, their commutator is defined to be $[\mathcal{A},\mathcal{B}]\triangleq 
\mathcal{AB}-\mathcal{BA}$. The Kronecker delta function is denoted by $\delta _{jk}$, whereas the Dirac delta function is denoted by $\delta (t)$.
The $m$-fold integral $\int_{-\infty }^{\infty }\cdots \int_{-\infty
}^{\infty }dt_{1}\cdots dt_{m}$ is sometimes denoted by $\int d\overrightarrow{t}$. Given a function $f(t)$ in the time domain, define its 
\textit{two-sided} Laplace transform \cite[Eq. (13)]{Sog10} to be $F[s]\equiv \mathscr{L}_{b}\{f(t)\}(s)\triangleq \int_{-\infty }^{\infty
}e^{-st}f(t)dt$. The $m$ dimensional Fourier transform of an $m$-variable function $f(t_1,\ldots,t_m)$ is, \cite{Bracewell99}, 
\begin{equation}\label{eq:fourier}
f(\mathrm{i}\omega _{1},\ldots ,\mathrm{i}\omega _{m})
\triangleq
 \frac{1}{(2\pi )^{m/2}}\int_{-\infty }^{\infty }\cdots \int_{-\infty }^{\infty
}dt_{1}\cdots dt_{m}\ e^{-\mathrm{i}(\omega _{1}t_{1}+\cdots +\omega
_{m}t_{m})}f(t_{1},\ldots ,t_{m}). 
 \end{equation}
We set $\hbar =1$ throughout this paper.


\section{Preliminaries}\label{sec:prelim}

In this section, quantum linear systems are briefly introduced; more
discussions can be found in, e.g., \cite{GZ00}, \cite{YK03a}, \cite{YK03b}, 
\cite{WM08}, \cite{GJ09}, \cite{WM10}, \cite{JG10}, \cite{IRP11}, \cite%
{zhang11}, \cite{zhang12}, \cite{NY13}, \cite{HIN14}, and \cite{zhang16}.
Some tensors and their associated operations are also discussed.

\subsection{Quantum linear systems}\label{sec:systems}

\begin{figure}[ptb]
\centering
\includegraphics[width=2.0in]{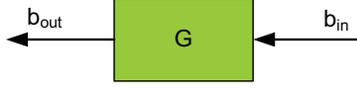}
\caption{Quantum linear system $G$ with input $b_{\mathrm{in}}$ and output $
b_{\mathrm{out}}$}
\label{system}
\end{figure}

A quantum linear system $G$ is shown schematically in Fig.~\ref{system}. In
this model, the quantum linear system $G$ consists of a collection of $n$
(interacting) quantum harmonic oscillators represented by $a=[a_{1} \ \cdots
\ a_{n}]^{T}$. Here, $a_{j}$ ($j=1,\ldots ,n$), defined on a Hilbert space $
\mathfrak{H}$, is the annihilation operator of the $j$th quantum harmonic
oscillator. The adjoint operator of $a_{j}$, denoted by $a_{j}^{\ast }$, is
called a creation operator. These operators satisfy the following canonical
commutation relations: $[a_{j},a_{k}^{\ast }]=\delta _{jk}$, and $
[a_{j},a_{k}]=[a_{j}^{\ast },a_{k}^{\ast }]=0$, $(j,k=1,\ldots ,n)$. The
input fields are represented by a vector of annihilation operators $b_{
\mathrm{in}}(t)=[b_{\mathrm{in},1}(t)\ \cdots \ b_{\mathrm{in},m}(t)]^{T}$;
the entry $b_{\mathrm{in},j}(t)$ ($j=1,\ldots ,m$), defined on a Fock space $
\mathfrak{F}$, is the annihilation operator for the $j$th input channel. The
adjoint operator of $b_{\mathrm{in},j}(t)$, denoted by $b_{\mathrm{in}
,j}^{\ast }(t)$, is also called a creation operator. However, unlike $a_{j}$
and $a_{k}^{\ast }$, the annihilation and creation operators for the input
fields satisfy the following singular commutation relations, \cite{GZ00}, 
\cite[Eq. (20)]{GJN10}, 
\begin{equation}
\lbrack b_{\mathrm{in},j}(t),b_{\mathrm{in},k}^{\ast }(r)] =\delta
_{jk}\delta (t-r), \ \ \lbrack b_{\mathrm{in},j}(t),b_{\mathrm{in},k}(r)]
=[b_{\mathrm{in},j}^{\ast }(t),b_{\mathrm{in},k}^{\ast }(r)]=0,\
j,k=1,\ldots ,m,~\forall t,r\in \mathbb{R}.  \label{eq:ccr}
\end{equation}
Notice the presence of the Dirac delta function $\delta (t-r)$ in Eq. (\ref
{eq:ccr}). Mathematically, it is often more convenient to work with
integrated annihilation and creation operators, which are defined
respectively to be $B_{\mathrm{in}}(t)\triangleq \int_{t_{0}}^{t}b_{\mathrm{
in}}(\tau )d\tau $ and $B_{\mathrm{in}}^{\#}(t)\triangleq \int_{t_{0}}^{t}b_{
\mathrm{in}}^{\#}(\tau )d\tau $, where the lower limit $t_{0}$ of the
integrals is the initial time, namely the time when the system and the
fields start to interact. The input gauge process (also called {\it number
} process) is defined by the following $m$-by-$m$ matrix of operators, \cite[
Chapter 11]{GZ00}, \cite[Section III.A]{GJN10}, \cite[Eq. (11)]{zhang13},
\begin{equation}
\Lambda _{\mathrm{in}}(t)\triangleq \int_{t_{0}}^{t}b_{\mathrm{in}
}^{\#}(\tau )b_{\mathrm{in}}^{T}(\tau )d\tau .  \label{july20_1}
\end{equation} 
In this paper, we deal with \emph{canonical }quantum input fields, that is,
the only non-zero It\^{o} products for the input fields are, \cite[Chapter 11]{GZ00}, \cite{GJ09}
, \cite{GJN10}, \cite[Eq. (12)]{zhang13}, 
\begin{align}
&
dB_{\mathrm{in},j}(t)dB_{\mathrm{in},k}^{\ast }(t) =\delta
_{jk}dt,~d\Lambda _{\mathrm{in}}^{jk}(t)dB_{\mathrm{in},l}^{\ast }(t)=\delta
_{kl}dB_{\mathrm{in},j}^{\ast }(t), 
 \nonumber
  \\
&
 dB_{\mathrm{in},j}(t)d\Lambda _{\mathrm{in}}^{kl}(t) =\delta _{jk}dB_{
\mathrm{in},l}(t),~d\Lambda _{\mathrm{in}}^{jk}(t)d\Lambda _{\mathrm{in}
}^{lr}(t)=\delta _{kl}d\Lambda _{\mathrm{in}}^{jr}(t),~j,k,l,r=1,\ldots
,m,~\forall t\in \mathbb{R}, 
 \label{Eq:CCR2}
\end{align}
where $\Lambda _{\mathrm{in}}^{jk}(t)$ is the entry of the matrix $\Lambda _{\mathrm{in}}(t)$ on the $j$th row and $k$th column, as introduced in the {\it Notation} part.

The dynamics of the open quantum linear system $G$ can be described
conveniently in the $(S_{-},L,H)$ formalism \cite{GJ09}, \cite{zhang12}.
Here, $S_{-}$ is a constant unitary matrix of dimension $m$, which can be
used to model static devices such as phase shifters and
beamsplitters. The operator $L$ describes how the system is coupled to the
fields, and is of the form $L=C_{-}a+C_{+}a^{\#}$ with $C_{-},C_{+}\in 
\mathbb{C}^{m\times n}$. For example, when a single-mode (namely, $n=1$) optical cavity is driven by a
light field, $L$ can be of the form $L=\sqrt{\kappa }a$, where $a$ is the
annihilation operator of the quantum harmonic operator for the cavity
(also called the cavity mode) and $\kappa >0$ is the coupling strength between
the cavity and the field. The operator $H$ stands for the initial system
Hamiltonian, which can be written as $H=\frac{1}{2}\breve{a}^{\dag }\Delta
\left( \Omega _{-},\Omega _{+}\right) \breve{a}$ with constant matrices $
\Omega _{-},\Omega _{+}\in \mathbb{C}^{n\times n}$ satisfying $\Omega
_{-}=\Omega _{-}^{\dag }$ and $\Omega _{+}=\Omega _{+}^{T}$. For example,
for the optical cavity above mentioned, $H=\frac{1}{2 } [ a^\ast \ \ a] 
\left[
\begin{smallmatrix}
\omega _{d} & 0 \\ 
0 & \omega _{d}
\end{smallmatrix}
\right] [a \ \ a^\ast]^T = \omega _{d}a^{\ast }a+\frac{1}{2}\omega _{d} $,
where $\omega _{d} \in \mbb{R}$ is the detuning frequency between the cavity mode and
the center frequency of the input light field. (The term $\frac{1}{2}\omega
_{d}$ introduces a global phase shift and leads to no consequence.) With
these parameters, in It\^{o} form, Schr\"{o}dinger's equation for the
temporal evolution of the open quantum linear system in Fig.~\ref{system}
is, \cite{HP84}, \cite[Eq. (30)]{GJ09}, \cite[Eq. (22)]{GJN10}, \cite[Eq.
(13)]{zhang13}, 
\begin{equation}  \label{eq:U}
dU(t, t_{0})=\left\{ \mathrm{Tr}[(S_{-}-I_{m})d\Lambda _{\mathrm{in}
}(t)^{T}]+dB_{\mathrm{in}}^{\dag }(t)L-L^{\dag }S_{-}dB_{\mathrm{in}}(t)-(
\frac{1}{2}L^{\dag }L+\mathrm{i}H)dt\right\} U(t, t_{0}),~t\geq t_{0}
\end{equation}
with  $U(t, t_{0})=I$ (identity operator) for all $t\leq t_{0}$.

In the Heisenberg picture, system operators evolve according to $\breve{a}
(t)=U(t, t_{0})^{\ast }\breve{a}(t_{0})U(t, t_{0})$ (component-wise for the
components of $\breve{a}(t_{0})$). Moreover, the output field $b_{\mathrm{out}}(t)$
carries away information of the system after interaction, and is defined by 
\begin{equation}  \label{eq:b_out_dec22}
\breve{b}_{\mathrm{out}}(t)\triangleq U(t, t_{0})^{\ast }\breve{b}_{\mathrm{in
}}(t)U(t, t_{0})
\end{equation}
(component-wise for the components of $\breve{b}_{\mathrm{in}}(t)$).
Consequently, by Eq. (\ref{eq:U}) and quantum It\^{o} calculus \cite{HP84},
Heisenberg's equation of motion for the system in Fig.~\ref{system} is, \cite
[ Eq. (26)]{GJN10}, \cite[Eqs. (14)-(15)]{zhang13}, 
\begin{align}
\dot{\breve{a}}(t)& =\mathbf{A}\breve{a}(t)+\mathbf{B}\breve{b}_{\mathrm{in}
}(t),  
\nonumber \\
\breve{b}_{\mathrm{out}}(t)& =\mathbf{C}\breve{a}(t)+\mathbf{S}\breve{b}_{
\mathrm{in}}(t),\ \ \breve{a}(t_{0})=\breve{a},  \label{system-out}
\end{align}
in which the constant system matrices are 
\begin{equation}
\mathbf{S}=\Delta (S_{-},0),\ \mathbf{C}=\Delta (C_{-},C_{+}),\ \mathbf{B}=-
\mathbf{C}^{\flat }\mathbf{S},\ \mathbf{A}=-\frac{1}{2}\mathbf{C}^{\flat }
\mathbf{C}-\mathrm{i}J_{n}\Delta (\Omega _{-},\Omega _{+}) 
 \label{eq:ABCD}
\end{equation}
with the matrix $J_{n}=\mathrm{diag}(I_{n},-I_{n})$ introduced in the \textit{
Notation} part. The gauge process $\Lambda _{\mathrm{out}}(t)$ of the output
fields, 
\begin{equation}
\Lambda _{\mathrm{out}}(t)\triangleq \int_{t_{0}}^{t}b_{\mathrm{out}
}^{\#}(\tau )b_{\mathrm{out}}^{T}(\tau )d\tau =U(t, t_{0})^{\ast }\Lambda _{
\mathrm{in}}(t)U(t, t_{0}),  \label{output_gauge}
\end{equation}   
satisfies the following quantum stochastic differential equation (QSDE), 
\cite{GJ09}, \cite[Eq. (16)]{zhang12}, 
\begin{equation}
d\Lambda _{\mathrm{out}}(t)=S_{-}^{\#}d\Lambda _{\mathrm{in}
}(t)S_{-}^{T}+S_{-}^{\#}dB_{\mathrm{in}}^{\#}(t)L^{T}(t)+L^{\#}(t)dB_{
\mathrm{in}}^{T}(t)S_{-}^{T}+L^{\#}(t)L^{T}(t)dt.  \label{eq:output_gauge}
\end{equation}
In quantum optics, the diagonal elements of $\Lambda _{\mathrm{out}}(t)$ are operators for the
total number of photons in each of the $m$ output channels, counted from
time $t_{0}$ to $t$. The intensity of the output field, namely the rate of
change of the number process $\Lambda _{\mathrm{out}}(t)$, is given by, \cite
[Eq. (45)]{zhang13}, 
\begin{equation}
\bar{n}_{\mathrm{out}}(t)\triangleq \langle \phi \Psi _{\mathrm{in}}|b_{
\mathrm{out}}^{\#}(t)b_{\mathrm{out}}^{T}(t)|\phi \Psi _{\mathrm{in}}\rangle
.  \label{n_out2}
\end{equation}
In Eq. (\ref{n_out2}), $|\phi \rangle $ is the initial system state and $
|\Psi _{\mathrm{in}}\rangle $ is the initial input field state. Therefore, the ket vector $ |\phi\rangle \otimes |\Psi _{\mathrm{in}}\rangle \equiv |\phi \Psi _{\mathrm{in}}\rangle $
is the initial joint system-field state. The bra vector $\langle \phi \Psi _{
\mathrm{in}}|$ is the Hilbert space conjugate of the ket vector $|\phi \Psi
_{\mathrm{in}}\rangle $. In this paper, $|\phi \rangle $ is always assumed
to be the vacuum state, while the specific form of $|\Psi _{\mathrm{in}
}\rangle $ will be given in due course.

The quantum linear system $G$ is said to be \textit{asymptotically stable}
if the matrix $\mathbf{A}$ in Eq. (\ref{eq:ABCD}) is Hurwitz stable, \cite[Sec.~III-A]{zhang11}. In
analogy to classical (namely non-quantum) control theory, the impulse
response function of the system $G$ is, \cite[Eq. (18)]{zhang13}, 
\begin{equation*}
g_{G}(t)\triangleq \left\{ 
\begin{array}{ll}
\delta (t)\mathbf{S}-\mathbf{C}e^{\mathbf{A}t}\mathbf{C}^{\flat }\mathbf{S},
& t\geq 0, \\ 
0, & t<0,
\end{array}
\right.  \label{eq:gg}
\end{equation*}
which enjoys the following doubled-up form 
\begin{equation}
g_{G}(t)=\Delta ( g_{G^{-}}(t),g_{G^{+}}(t)) ,  \label{eq:impulse}
\end{equation}
with matrix functions 
\begin{align}
g_{G^{-}}(t)& \triangleq \left\{ 
\begin{array}{ll}
\delta (t)S_{-}-[
\begin{array}{cc}
C_{-} & C_{+}
\end{array}
]e^{\mathbf{A}t}\left[ 
\begin{array}{c}
C_{-}^{\dag } \\ 
-C_{+}^{\dag }
\end{array}
\right] S_{-}, & t\geq 0, \\ 
0, & t<0,
\end{array}
\right.  \nonumber \\
g_{G^{+}}(t)& \triangleq \left\{ 
\begin{array}{ll}
-[
\begin{array}{cc}
C_{-} & C_{+}
\end{array}
]e^{\mathbf{A}t}\left[ 
\begin{array}{c}
-C_{+}^{T} \\ 
C_{-}^{T}
\end{array}
\right] S_{-}^{\#}, & t\geq 0, \\ 
0, & t<0.
\end{array}
\right.  \label{eq:io}
\end{align}
Next, we express the output field in terms of the impulse function $g_G(t)$. In fact, solving Eq. (\ref{system-out}) we have 
\begin{equation}
\breve{b}_{\mathrm{out}}(t)=\mathbf{C}e^{\mathbf{A}(t-t_{0})}\breve{a}
+\int_{t_{0}}^{t}g_{G}(t-r)\breve{b}_{\mathrm{in}}(r)dr.  
\label{eq:out_tf}
\end{equation}
Furthermore, if the system is asymptotically stable, then in the limit $t_{0}\rightarrow
-\infty $, Eq. (\ref{eq:out_tf}) reduces to
\begin{equation}
\breve{b}_{\mathrm{out}}(t)
=
\int_{-\infty }^{t}g_{G}(t-r)\breve{b}_{\mathrm{
in}}(r)dr = g_{G}\circledast \breve{b}_{\mathrm{in}}(t). 
 \label{eq:out_tf_2}
\end{equation}


\begin{remark}
If the interaction starts in the remote past, namely $t_{0}\rightarrow
-\infty $, and if the system is asymptotically stable, Eq. (\ref{eq:out_tf_2}) indicates that the initial system information has no influence on the
output field. This is also true in classical control theory, see, e.g., \cite{KS72}.
\end{remark}

Define a matrix function 
\begin{equation}
g_{G^{-1}}(t)\triangleq \Delta (g_{G^{-}}(-t)^{\dagger
},-g_{G^{+}}(-t)^{T}) .  \label{eq:G_inv}
\end{equation}
It can be verified that the following convolution relations 
\begin{equation}
g_{G}\circledast g_{G^{-1}}\circledast f(t)=g_{G^{-1}}\circledast
g_{G}\circledast f(t)=f(t)  \label{eq:may9_1}
\end{equation}
hold for any function $f(t)$ of suitable dimension provided that the involved integrals converge. Thus, $g_{G^{-1}}(t)$
is the inverse function of the impulse response function $g_{G}(t)$.
According to Eqs. (\ref{eq:out_tf_2}) and (\ref{eq:may9_1}), in the limit $
t_{0}\rightarrow -\infty $ we have
\begin{equation}
\breve{b}_{\mathrm{in}}(t)=g_{G^{-1}}\circledast \breve{b}_{\mathrm{out}}(t).
\label{eq:out_tf_3}
\end{equation}

A class of \textit{passive} quantum linear systems is obtained when $C_{+}=0$
and $\Omega _{+}=0$ in Eq. (\ref{eq:ABCD}). For this type of systems, it is
sufficient to work in the annihilation-operator representation. To be
specific, it suffices to study 
\begin{eqnarray}
\dot{a}(t) &=&Aa(t)+Bb_{\mathrm{in}}(t),  \nonumber \\
b_{\mathrm{out}}(t) &=&Ca(t)+S_{-}b_{\mathrm{in}}(t),~~~a(t_{0})=a,
\label{sys_passive}
\end{eqnarray}
where 
\begin{equation*}
A=-\mathrm{i}\Omega _{-}-\frac{1}{2}C_{-}^{\dag }C_{-},~~~B=-C_{-}^{\dag
}S_{-},~~~C=C_{-}.
\end{equation*}
In this case, Eq. (\ref{eq:io}) reduces to
\begin{equation}
g_{G^{-}}(t)=\left\{ 
\begin{array}{ll}
\delta (t)S_{-}-C_{-}e^{At}C_{-}^{\dag }S_{-}, & t\geq 0, \\ 
0, & t<0,
\end{array}
\right. \ \ \ \ \ \  g_{G^{+}}(t)=0.  \label{july17_1}
\end{equation}
Accordingly, Eqs.  (\ref{eq:impulse}) and (\ref{eq:G_inv}) reduce to 
\begin{equation}
g_{G}(t)=\Delta \left( g_{G^{-}}(t),0\right) ,   \ \ 
g_{G^{-1}}(t)\triangleq \Delta \left( g_{G^{-}}(-t)^{\dagger },0\right) ,
\label{dec1_g_inv}
\end{equation}
respectively.

It is well-known that in linear classical control theory,  if an asymptotically stable finite-dimensional linear time-invariant (FDLTI) system is driven by Gaussian white noise, then the steady-state output is again a Gaussian stationary process, see, e.g., \cite[Section 11, Chapter 1]{KS72}, \cite{AM79}, and  \cite{BH97}.  The following result is the quantum counterpart.  

\begin{lemma}
\label{lem:gaussian}\cite[Theorem 2]{zhang13} Let the asymptotically stable
quantum linear system $G$ be initialized in the vacuum state $|\phi \rangle $
and let the input field $|\Psi _{\mathrm{in}}\rangle $ be in the vacuum
state $|0\rangle $. Then the steady-state 
output field state is a zero-mean Gaussian state, whose power spectral density
matrix $R_{\mathrm{out}}[\mathrm{i}\omega ]$ is given by 
\begin{equation}
R_{\mathrm{out}}[\mathrm{i}\omega ]=G[\mathrm{i}\omega ]\left[ 
\begin{array}{cc}
I_{m} & 0 \\ 
0 & 0_{m}
\end{array}
\right] G[\mathrm{i}\omega ]^{\dag },  \label{eq:temp_gaussian}
\end{equation}
where $G[\mathrm{i}\omega ]=\int_{-\infty }^{\infty }e^{-\mathrm{i}\omega
t}g_{G}(t)dt$ is the two-sided Laplace transform of $g_{G}(t)$ with $s=\mathrm{i}\omega $, as
introduced in the Notation part. In particular, if the system is passive,
then the output is in a vacuum state with power spectral density matrix 
\begin{equation*}
R_{\mathrm{out}}[\mathrm{i}\omega ]=\left[ 
\begin{array}{cc}
I_{m} & 0 \\ 
0 & 0_{m}
\end{array}
\right] .  \label{R_out_vac}
\end{equation*}
\end{lemma}

\subsection{Tensors}\label{subsec:tensor}

Tensors and their associated operations are essential
mathematical machinery for the research carried out in this paper \cite
{QSW07}, \cite{KB09}, \cite{zhang14}. In this subsection, we discuss several
tensors.

Given an $m$-variable function $\psi (t_{1},\ldots ,t_{m})$ and an $m$-dimensional column vector $x(t)=[x_{1}(t)\ \cdots \ x_{m}(t)]^{T}$, denote
\begin{equation}
\psi \circ ^{m}x\equiv \int_{-\infty }^{\infty }\cdots \int_{-\infty
}^{\infty }dt_{1}\cdots dt_{m}\ \psi (t_{1},\ldots ,t_{m})x_{1}(t_{1})\cdots
x_{m}(t_{m}).  \label{dec3_circ}
\end{equation}
Given an $m$-way $m$-dimensional tensor function $\varphi =\left( \varphi
_{j_{1}\ldots j_{m}}(t_{1},\ldots ,t_{m})\right) $, ($j_{1},\ldots
,j_{m}=1,\ldots ,m$), and an $m$-dimensional column vector $
x(t)=[x_{1}(t)\ \cdots \ x_{m}(t)]^{T}$, denote 
\begin{equation}
\varphi \circledcirc ^{m}x
\equiv
 \sum_{j_{1}=1}^{m}\cdots \sum_{j_{m}=1}^{m}\int_{-\infty }^{\infty }\cdots \int_{-\infty }^{\infty
}dt_{1}\cdots dt_{m}\ \varphi _{j_{1}\ldots j_{m}}(t_{1},\ldots
,t_{m})x_{j_{1}}(t_{1})\cdots x_{j_{m}}(t_{m}). 
 \label{dec3_circ^2}
\end{equation}
We may update an $m$-variable function $\psi (t_{1},\ldots ,t_{m})$ to an $m$
-way $m$-dimensional tensor function $\psi ^{\uparrow }=( \psi^{\uparrow }
_{j_{1}\ldots j_{m}}(t_{1},\ldots ,t_{m})) $ with entries 
\begin{equation}
\psi _{j_{1}\ldots j_{m}}^{\uparrow }(t_{1},\ldots ,t_{m})\triangleq \left\{ 
\begin{array}{ll}
\psi (t_{1},\ldots ,t_{m}), & \mathrm{if\ }j_{1}=1,j_{2}=2,\ldots ,j_{m}=m,
\\ 
0, & \mathrm{otherwise}.
\end{array}
\right.  \label{dec_up1}
\end{equation}
Then Eq. (\ref{dec3_circ}) can be re-written as Eq. (\ref{dec3_circ^2}),
specifically,
\begin{equation*}
\psi \circ ^{m}x=\psi ^{\uparrow }\circledcirc ^{m}x.
\end{equation*}
Let\ $\psi (t_{1}^{1},\ldots ,t_{k_{1}}^{1},\ldots ,t_{1}^{m},\ldots
,t_{k_{m}}^{m})$ be an $N$-variable function, where the positive integers
$k_{1},\ldots ,k_{m}$ satisfy $\sum_{i=1}^{m}k_{i}=N$. Let $
x(t)=[x_{1}(t)\ \cdots \ x_{m}(t)]^{T}$ be an $m$-dimensional column vector.  Denote 
\begin{eqnarray}
&&
\psi \ \ \cdot _{k_{1}\cdots k_{m}}^{N}\ \ x  
\label{dec3_dot} \\
&\equiv &
\int_{-\infty }^{\infty }\cdots \int_{-\infty }^{\infty
}dt_{1}^{1}\cdots dt_{k_{1}}^{1}\cdots dt_{1}^{m}\cdots dt_{k_{m}}^{m}\ \psi
(t_{1}^{1},\ldots ,t_{k_{1}}^{1},\ldots ,t_{1}^{m},\ldots
,t_{k_{m}}^{m})x_{1}(t_{1}^{1})\cdots x_{1}(t_{k_{1}}^{1})\cdots
x_{m}(t_{1}^{m})\cdots x_{m}(t_{k_{m}}^{m}).  
\nonumber
\end{eqnarray}

Given an $N$-way $m$-dimensional tensor function $\varphi =\left( \varphi
_{j_{1}^{1}\ldots j_{k_{1}}^{1}\ldots j_{1}^{m}\ldots
j_{k_{m}}^{m}}(t_{1}^{1},\ldots ,t_{k_{1}}^{1},\ldots ,t_{1}^{m},\ldots
,t_{k_{m}}^{m})\right) $ and an $m$-dimensional vector $
x(t)=[x_{1}(t)\ \cdots \ x_{m}(t)]^{T}$, denote
\begin{eqnarray}
&&
\varphi  \odot _{k_{1}\cdots k_{m}}^{N}x  
\label{dec3_dot^2} 
\\
&\equiv & \sum_{j_{1}^{1}=1}^{m}\cdots \sum_{j_{k_{1}}^{1}=1}^{m}\cdots
\sum_{j_{1}^{m}=1}^{m}\cdots \sum_{j_{k_{m}}^{m}=1}^{m}\int_{-\infty
}^{\infty }\cdots \int_{-\infty }^{\infty }dt_{1}^{1}\cdots
dt_{k_{1}}^{1}\cdots dt_{1}^{m}\cdots dt_{k_{m}}^{m}\   \nonumber \\
&& \ \ \ \times \varphi _{j_{1}^{1}\ldots j_{k_{1}}^{1}\ldots
j_{1}^{m}\ldots j_{k_{m}}^{m}}(t_{1}^{1},\ldots ,t_{k_{1}}^{1},\ldots
,t_{1}^{m},\ldots ,t_{k_{m}}^{m})x_{j_{1}^{1}}(t_{1}^{1})\cdots
x_{j_{k_{1}}^{1}}(t_{k_{1}}^{1})\cdots x_{j_{1}^{m}}(t_{1}^{m})\cdots
x_{j_{k_{m}}^{m}}(t_{k_{m}}^{m}).  \nonumber
\end{eqnarray}
Update the $N$-variable function $\psi (t_{1}^{1},\ldots
,t_{k_{1}}^{1},\ldots ,t_{1}^{m},\ldots ,t_{k_{m}}^{m})$ in Eq. (\ref
{dec3_dot}) to an $N$-way $m$-dimensional tensor function $\psi ^{\uparrow }(t_{1}^{1},\ldots
,t_{k_{1}}^{1},\ldots ,t_{1}^{m},\ldots ,t_{k_{m}}^{m})$, whose elements are
defined as 
\begin{eqnarray}
&&\psi _{j_{1}^{1}\ldots j_{k_{1}}^{1}\ldots j_{1}^{m}\ldots
j_{k_{m}}^{m}}^{\uparrow }(t_{1}^{1},\ldots ,t_{k_{1}}^{1},\ldots
,t_{1}^{m},\ldots ,t_{k_{m}}^{m})  \nonumber \\
&\triangleq &\left\{ 
\begin{array}{ll}
\psi (t_{1}^{1},\ldots ,t_{k_{1}}^{1},\ldots ,t_{1}^{m},\ldots
,t_{k_{m}}^{m}), & \mathrm{if\ }j_{1}^{1}=1,\cdots
,j_{k_{1}}^{1}=k_{1},\cdots ,j_{1}^{m}=\sum\limits_{j=1}^{m-1}k_{j}+1,\cdots
,j_{k_{m}}^{m}=N, \\ 
0, & \mathrm{otherwise}.
\end{array}
\right.   \label{dec_up2}
\end{eqnarray}
Then
\begin{equation*}
\psi \ \ \cdot _{k_{1}\cdots k_{m}}^{N}\ \ x=\psi ^{\uparrow }\odot
_{k_{1}\cdots k_{m}}^{N}x.
\end{equation*}

In the above, we have defined several operations between tensors 
and vectors.  In the following, we look at operations between tensors and matrices.

Given an $m\times m$ matrix function $\mathcal{A}(t)$ and an $m$-way $m$-dimensional tensor function $\varphi =\left( \varphi _{j_{1}\ldots
j_{m}}(t_{1},\ldots ,t_{m})\right) $, $(j_{1},\ldots ,j_{m}=1,\ldots ,m)$,
define another $m$-way $m$-dimensional tensor function $\tilde{\varphi}=(
\tilde{\varphi}_{i_{1}\ldots i_{m}}(r_{1},\ldots ,r_{m}))$ in such a way
that, for all $i_{1},\ldots ,i_{m}=1,\ldots ,m$, 
\begin{equation}
\tilde{\varphi}_{i_{1}\ldots i_{m}}(r_{1},\ldots ,r_{m})=\sum_{j_{1},\ldots
,j_{m}=1}^{m}\int_{-\infty }^{\infty }\cdots \int_{-\infty }^{\infty
}dt_{1}\cdots dt_{m}\ \mathcal{A}^{i_{1}j_{1}}(r_{1}-t_{1})\cdots \mathcal{A}
^{i_{m}j_{m}}(r_{m}-t_{m})\varphi _{j_{1}\ldots j_{m}}(t_{1},\ldots ,t_{m}).
\label{eq:product_m_a}
\end{equation}
Eq. (\ref{eq:product_m_a}) may be re-written in a more compact form 
\begin{equation}
\tilde{\varphi}=\varphi \circledast _{t}^{m}\mathcal{A},
\label{eq:product_m_b}
\end{equation}
where the subscript ``$t$'' indicates the
time domain, while the superscript ``$m$''
implies the $m$-fold convolution. Applying the $m$-dimensional Fourier
transform (\ref{eq:fourier}) to Eq. (\ref{eq:product_m_a}), we get
\begin{equation}
\tilde{\varphi}_{i_{1}\ldots i_{m}}(\mathrm{i}\omega _{1},\ldots ,\mathrm{i}
\omega _{m})=\sum_{j_{1},\ldots ,j_{m}=1}^{m}\mathcal{A}^{i_{1}j_{1}}[
\mathrm{i}\omega _{1}]\cdots \mathcal{A}^{i_{m}j_{m}}[\mathrm{i}\omega
_{m}]\varphi _{j_{1}\ldots j_{m}}(\mathrm{i}\omega _{1},\ldots ,\mathrm{i}
\omega _{m}),  \label{eq:product_m_c}
\end{equation}
where
\begin{equation*}
\mathcal{A}^{i_{k}j_{k}}[\mathrm{i}\omega _{k}]=\int_{-\infty }^{\infty }e^{-
\mathrm{i}\omega _{k}t}\mathcal{A}^{i_{k}j_{k}}(t)dt
\end{equation*}
is the two-sided Laplace transform of $\mathcal{A}^{i_{k}j_{k}}(t)$. In
analogy to Eq. (\ref{eq:product_m_b}), we may also write Eq. (\ref
{eq:product_m_c}) in the following compact form 
\begin{equation*}
\tilde{\varphi}=\varphi \circledast _{\omega }^{m}\mathcal{A},
\label{eq:product_m_d}
\end{equation*}
where the subscript ``$\omega $'' indicates
the frequency domain.

Given an $N$-way $m$-dimensional tensor function $\varphi =\left( \varphi
_{j_{1}^{1}\ldots j_{k_{1}}^{1}\ldots j_{1}^{m}\ldots
j_{k_{m}}^{m}}(t_{1}^{1},\ldots ,t_{k_{1}}^{1},\ldots ,t_{1}^{m},\ldots
,t_{k_{m}}^{m})\right) $, and an $m\times m$ matrix function $\mathcal{A}(t)$%
, define a new $N$-way $m$-dimensional tensor function $\tilde{\varphi}$ by 
\begin{eqnarray*}
&&\tilde{\varphi}_{l_{1}^{1}\ldots l_{k_{1}}^{1}\ldots l_{1}^{m}\ldots
l_{k_{m}}^{m}}(r_{1}^{1},\ldots ,r_{k_{1}}^{1},\ldots ,r_{1}^{m},\ldots
,r_{k_{m}}^{m})  
\label{dec11_1} \\
&\triangleq &\sum_{i_{1}^{1}=1}^{m}\cdots \sum_{i_{k_{1}}^{1}=1}^{m}\cdots
\sum_{i_{1}^{m}=1}^{m}\cdots \sum_{i_{k_{m}}^{m}=1}^{m}\int_{-\infty
}^{\infty }\cdots \int_{-\infty }^{\infty }dt_{1}^{1}\cdots
dt_{k_{1}}^{1}\cdots dt_{1}^{m}\cdots dt_{k_{m}}^{m}\ \mathcal{A}
^{l_{1}^{1}i_{1}^{1}}(r_{1}^{1}-t_{1}^{1})\cdots \mathcal{A}
^{l_{k_{1}}^{1}i_{k_{1}}^{1}}(r_{k_{1}}^{1}-t_{k_{1}}^{1})\times \cdots 
\nonumber \\
&&\;\times \mathcal{A}^{l_{1}^{m}i_{1}^{m}}(r_{1}^{m}-t_{1}^{m})\cdots 
\mathcal{A}^{l_{k_{m}}^{m}i_{k_{m}}^{m}}(r_{k_{m}}^{m}-t_{k_{m}}^{m})\varphi
_{i_{1}^{1}\ldots i_{k_{1}}^{1}\ldots i_{1}^{m}\ldots
i_{k_{m}}^{m}}(t_{1}^{1},\ldots ,t_{k_{1}}^{1},\ldots ,t_{1}^{m},\ldots
,t_{k_{m}}^{m}),  \nonumber
\end{eqnarray*}
which may be re-written in a more compact form 
\begin{equation}
\tilde{\varphi}=\varphi \circledast _{t, k_{1}\cdots k_{m}}^{N}\mathcal{A}.
\label{dec11_2}
\end{equation}

Given an $m$-way $m$-dimensional tensor function $\varphi = \varphi _{j_1
\dots j_m}(\mathrm{i}\omega _{1},\ldots ,\mathrm{i}\omega _{m})$ in the frequency domain, denote
\begin{equation*}
\left\Vert \varphi (\mathrm{i}\omega _{1},\ldots ,\mathrm{i}\omega
_{m})\right\Vert \equiv \sqrt{\sum_{j_{1},\ldots ,j_{m}=1}^{m}\left\vert
\varphi _{j_{1}\ldots j_{m}}(\mathrm{i}\omega _{1},\ldots ,\mathrm{i}\omega
_{m})\right\vert ^{2}},~~\forall \omega _{1},\ldots ,\omega _{m}\in \mathbb{R}.  \label{eq:norm2}
\end{equation*}
We end this subsection by citing the following result.

\begin{lemma}
\label{lem:norm} \cite[Theorem 3.3]{QSW07} Let two tensors $\varphi $ and $
\tilde{\varphi}$ be related by Eq. (\ref{eq:product_m_c}), or equivalently
Eq. (\ref{eq:product_m_a}). If $\mathcal{A}[\mathrm{i}\omega ]$ is unitary
for all $\omega \in \mathbb{R}$, then
\begin{equation*}
\left\Vert \tilde{\varphi}(\mathrm{i}\omega _{1},\ldots ,\mathrm{i}\omega
_{m})\right\Vert =\left\Vert \varphi (\mathrm{i}\omega _{1},\ldots ,\mathrm{i
}\omega _{m})\right\Vert ,~~\forall \omega _{1},\ldots ,\omega _{m}\in 
\mathbb{R}.  \label{eq:norm3}
\end{equation*}
\end{lemma}


\section{$m$ photons superposed among $m$ input channels}\label{sec:mm}

In this section, we investigate how a quantum linear system responds to a class of $m$-photon input states. We first define $m$-photon input states in Subsection \ref
{subsec:m_photon}, then derive the output intensity in Subsection \ref
{subsec:intensity}, after that, we present an analytic form of the output
field state when the underlying quantum linear system is passive in
Subsections \ref{sec:passive_mm} and \ref{subsec:invariant}, finally we turn
to the non-passive case in Subsection \ref{subsec:nonpassive}.


\subsection{$m$-photon input states}

\label{subsec:m_photon}

In this subsection, we introduce a class of $m$-photon input states. For
ease of presentation, we start with the \textit{single-channel
single-photon} state case. In this case, $m=1$. A single-channel
single-photon input state can be defined by 
\begin{equation*}
\left\vert \Psi _{\mathrm{in}}\right\rangle \triangleq \int_{-\infty
}^{\infty }dt\ \psi _{\mathrm{in}}(t)b_{\mathrm{in}}^{\ast }(t)\left\vert
0\right\rangle .  \label{eq:1_photon_t}
\end{equation*}
Here, the function $\psi _{\mathrm{in}}$ is square integrable, more
specifically, $\psi _{\mathrm{in}}\in L_{2}(\mbb{R},\mathbb{C})$. The  Euclidean norm of $\psi _{\mathrm{in}}$, $\Vert \psi _{\mathrm{in}}\Vert \triangleq 
\sqrt{\int_{-\infty }^{\infty }|\psi _{\mathrm{in}}(t)|^{2}dt}$, is equal to
1. Consequently, the inner product $\langle \Psi _{\mathrm{in}}|\Psi _{
\mathrm{in}}\rangle =1$. That is, $\left\vert \Psi _{\mathrm{in}
}\right\rangle $ is a normalized state. Moreover, it can be easily shown
that 
\begin{equation}
\lim_{t_{0}\rightarrow -\infty ,t\rightarrow \infty }\langle \Psi _{\mathrm{
in}}|\Lambda _{\mathrm{in}}(t)|\Psi _{\mathrm{in}}\rangle =1,  \label{aug1_1}
\end{equation}
where $\Lambda _{\mathrm{in}}(t)$ is the input gauge process defined in Eq.
(\ref{july20_1}). Eq. (\ref{aug1_1}) indicates that there is one
photon in the field. On the other hand, it can be readily verified that 
\begin{equation}
\langle \Psi _{\mathrm{in}}|b_{\mathrm{in}}(t)|\Psi _{\mathrm{in}}\rangle
=\langle \Psi _{\mathrm{in}}|b_{\mathrm{in}}^{\ast }(t)|\Psi _{\mathrm{in}
}\rangle =0,~\forall t\in \mathbb{R}.  \label{aug1_2}
\end{equation}
That is, the average field amplitude is zero. Finally, it is worth noting
that $|\Psi _{\mathrm{in}}\rangle $ is not a single-photon \textit{coherent}
state which can be defined to be 
\begin{equation*}
|\alpha _{\psi _{\mathrm{in}}}\rangle \triangleq \exp \left( \int_{-\infty
}^{\infty }dt\ \alpha \psi _{\mathrm{in}}(t)b_{\mathrm{in}}^{\ast
}(t)-\int_{-\infty }^{\infty }dt\ (\alpha \psi _{\mathrm{in}}(t))^{\ast }b_{
\mathrm{in}}(t)\right) |0\rangle ,  \label{coherent}
\end{equation*}
where $\alpha =e^{i\theta }$ is a complex number. In fact, for the
single-photon coherent state $|\alpha _{\psi _{\mathrm{in}}}\rangle $, Eq. (\ref{aug1_1}) still holds, but Eq. (\ref{aug1_2}) does not.

Next, we look at \textit{single-channel} \textit{two-photon} states, which
can be defined as 
\begin{equation}
|\Psi _{\mathrm{in}}\rangle \triangleq \int_{-\infty }^{\infty
}\int_{-\infty }^{\infty }dt_{1}dt_{2}\ \psi _{\mathrm{in}}(t_{1},t_{2})b_{
\mathrm{in}}^{\ast }(t_{1})b_{\mathrm{in}}^{\ast }(t_{2})|0\rangle .
\label{aug1_3}
\end{equation}
Here, the function $\psi _{\mathrm{in}}(t_{1},t_{2})$ is required to
normalize the state, namely $\langle \Psi _{\mathrm{in}}|\Psi _{\mathrm{in}
}\rangle =1$, which is equivalent to 
\begin{equation*}
2\int_{-\infty }^{\infty }\int_{-\infty }^{\infty }dt_{1}dt_{2}\ \left\vert
\psi _{\mathrm{in}}(t_{1},t_{2})\right\vert ^{2}=1.
\end{equation*}
Moreover, swapping $t_{1}$ and $t_{2}$ in Eq. (\ref{aug1_3}) yields 
\begin{equation}
|\Psi _{\mathrm{in}}\rangle =\int_{-\infty }^{\infty }\int_{-\infty
}^{\infty }dt_{1}dt_{2}\ \psi _{\mathrm{in}}(t_{2},t_{1})b_{\mathrm{in}
}^{\ast }(t_{1})b_{\mathrm{in}}^{\ast }(t_{2})|0\rangle . 
 \label{aug1_4}
\end{equation}
Comparing Eqs. (\ref{aug1_3}) and (\ref{aug1_4}) we see that $\psi _{\mathrm{
in}}(t_{1},t_{2})=\psi _{\mathrm{in}}(t_{2},t_{1})$. Moreover, it can be
verified that 
\begin{equation*}
\displaystyle{\lim_{t_{0}\rightarrow -\infty ,t\rightarrow \infty }}\langle
\Psi _{\mathrm{in}}|\Lambda _{\mathrm{in}}(t)|\Psi _{\mathrm{in}}\rangle =2,
\end{equation*}
i.e., there are two photons in the field. 

Next, let us look at 
\textit{two-channel} \textit{two-photon} states, which can be defined to be 
\begin{equation}
|\Psi _{\mathrm{in}}\rangle \triangleq \int_{-\infty }^{\infty
}\int_{-\infty }^{\infty }dt_{1}dt_{2}\ \psi _{\mathrm{in}}(t_{1},t_{2})b_{
\mathrm{in},1}^{\ast }(t_{1})b_{\mathrm{in},2}^{\ast }(t_{2})|0_{1}\rangle
\otimes |0_{2}\rangle .  \label{aug1_5}
\end{equation}
Again, the function $\psi _{\mathrm{in}}(t_{1},t_{2})$ is required to
normalize the state. This is guaranteed by 
\begin{equation}
\int_{-\infty }^{\infty }\int_{-\infty }^{\infty }dt_{1}dt_{2}\ \left\vert
\psi _{\mathrm{in}}(t_{1},t_{2})\right\vert ^{2}=1.  \label{aug1_7}
\end{equation}
(Notice that in this case, the condition $\psi _{\mathrm{in}
}(t_{1},t_{2})=\psi _{\mathrm{in}}(t_{2},t_{1})$ is not necessary.) It can
be easily shown that 
\begin{equation}
\lim_{t_{0}\rightarrow -\infty ,t\rightarrow \infty }\langle \Psi _{\mathrm{
in}}|\Lambda _{\mathrm{in}}(t)|\Psi _{\mathrm{in}}\rangle =\left[ 
\begin{array}{cc}
1 & 0 \\ 
0 & 1
\end{array}
\right] .  \label{aug1_18}
\end{equation}
Eq. (\ref{aug1_18}) implies that each channel contains one photon. However, these two photons can form an entangled state.  Moreover,
if we use the single-photon state $\int_{-\infty }^{\infty }\gamma (\tau )b_{
\mathrm{in},2}^{\ast }(\tau )d\tau \ |0_{2}\rangle $ to measure the second
channel, the resulting state for the first channel is
given by 
\begin{equation*}
\int_{-\infty }^{\infty }\left[ \int_{-\infty }^{\infty }\gamma ^{\ast
}(\tau )\psi _{\mathrm{in}}(t,\tau )d\tau \right] b_{\mathrm{in},1}^{\ast
} (t) dt\ |0_{1}\rangle .  \label{aug1_19}
\end{equation*}
In general, Eq. (\ref{aug1_5}) defines a state for which the two photons are
entangled. However, for the special case that $\psi _{\mathrm{in}
}(t_{1},t_{2})=\xi _{1}(t_{1})\xi _{2}(t_{2})$, we end up with a product
state 
\begin{equation}
|\Psi _{\mathrm{in}}\rangle =\int_{-\infty }^{\infty }\xi _{1}(t_{1})b_{
\mathrm{in},1}^{\ast }(t_{1})dt_{1}|0_{1}\rangle \otimes \int_{-\infty
}^{\infty }\xi _{2}(t_{2})b_{\mathrm{in},2}^{\ast
}(t_{2})dt_{2}|0_{2}\rangle .  \label{aug1_17}
\end{equation}
For the state defined in Eq. (\ref{aug1_17}), there exists no entanglement between these two photons.

We are ready to introduce a class of $m$-channel $m$-photon input states.
Such states can be of the form 
\begin{equation}
|\Psi _{\mathrm{in}}\rangle \triangleq \int_{-\infty }^{\infty }\cdots
\int_{-\infty }^{\infty }dt_{1}\cdots dt_{m}\ \psi _{\mathrm{in}
}(t_{1},\ldots ,t_{m})b_{\mathrm{in},1}^{\ast }(t_{1})\cdots b_{\mathrm{in}
,m}^{\ast }(t_{m})|0_{1}\rangle \otimes \cdots \otimes |0_{m}\rangle .
\label{eq:aug23_10}
\end{equation}
For convenience, in the sequel we use the shorthand notation $|0^{\otimes
m}\rangle $ for the tensor product of the vacuum input fields $|0_{1}\rangle
\otimes \cdots \otimes |0_{m}\rangle $. In the notation introduced in
Subsection \ref{subsec:tensor}, Eq. (\ref{eq:aug23_10}) may be re-written as 
\begin{equation*}  \label{dec23_1}
|\Psi _{\mathrm{in}}\rangle =\psi _{\mathrm{in}}\circ ^{m}b_{\mathrm{in}
}^{\#}\ |0^{\otimes m}\rangle .
\end{equation*}
By analogy with Eq. (\ref{aug1_7}), it can be readily shown that the
normalization condition for $|\Psi _{\mathrm{in}}\rangle $ is 
\begin{equation}
\left\Vert \psi _{\mathrm{in}}\right\Vert ^{2}\equiv \int_{-\infty }^{\infty
}\cdots \int_{-\infty }^{\infty }dt_{1}\cdots dt_{m}\ \left\vert \psi _{
\mathrm{in}}(t_{1},\ldots ,t_{m})\right\vert ^{2}=1.  \label{aug1_6}
\end{equation}
The bra vector $\left\langle \Psi _{\mathrm{in}}\right\vert $, namely the
conjugate of the ket vector $|\Psi _{\mathrm{in}}\rangle $, is 
\begin{equation}
\left\langle \Psi _{\mathrm{in}}\right\vert 
=
\langle 0^{\otimes m}\vert \int_{-\infty }^{\infty }\cdots
\int_{-\infty }^{\infty }dt_{1}\cdots dt_{m}\ \psi _{\mathrm{in}}^{\ast
}(t_{1},\ldots ,t_{m})b_{\mathrm{in},1}(t_{1})\cdots b_{\mathrm{in}
,m}(t_{m}).  
\label{aug22_1}
\end{equation}

For the $m$-channel $m$-photon state $|\Psi _{\mathrm{in}}\rangle $, it is
clear that
\begin{equation}
\left\langle \Psi _{\mathrm{in}}|\breve{b}_{\mathrm{in}}(t)|\Psi _{\mathrm{in
}}\right\rangle =0,~\forall t\in \mathbb{R}.  
\label{eq:amplitude}
\end{equation}
That is, the average field amplitude of the input light field is $0$. Next,
we look at two-time correlations $\langle \Psi _{\mathrm{in}}|\breve{b}_{\mathrm{in}}(t)\breve{b}_{\mathrm{in}}^{\dag
}(r)|\Psi _{\mathrm{in}}\rangle$ with $t,r\in \mbb{R}$. For each $k=1,\ldots ,m$, introduce the
notation 
\begin{equation}
\zeta _{k}(\tau ,r) \equiv \psi _{\mathrm{in}}(\tau _{1},\ldots ,\tau
_{k-1},r,\tau _{k+1},\ldots ,\tau _{m}).  
\label{eq:zeta_k}
\end{equation}
Specifically,
\begin{eqnarray*}
\zeta _{1}(\tau ,r) &=&\psi _{\mathrm{in}}(r,\tau _{2},\ldots ,\tau _{m}), \\
\zeta _{2}(\tau ,r) &=&\psi _{\mathrm{in}}(\tau _{1},r,\tau _{3},\ldots
,\tau _{m}), \\
&&\vdots \\
\zeta _{m}(\tau ,r) &=&\psi _{\mathrm{in}}(\tau _{1},\ldots ,\tau _{m-1},r).
\end{eqnarray*}
Also, define a diagonal matrix function
\begin{equation}
\Lambda \left( t, r\right) \triangleq \left[ 
\begin{array}{ccc}
\int_{-\infty }^{\infty }\cdots \int_{-\infty }^{\infty }d\tau _{2}\cdots
d\tau _{m}\ \zeta _{1}(\tau ,t)^{\ast }\zeta _{1}(\tau ,r) &  &  \\ 
& \ddots &  \\ 
&  & \int_{-\infty }^{\infty }\cdots \int_{-\infty }^{\infty }d\tau
_{1}\cdots d\tau _{m-1}\ \zeta _{m}(\tau ,t)^{\ast }\zeta _{m}(\tau ,r)
\end{array}
\right] ,\ \forall t, r\in \mathbb{R}.  
\label{eq:input_Lambda}
\end{equation}
Clearly, $\Lambda \left( t, r\right) ^\dag =\Lambda \left( r, t\right) $, and
by Eq. (\ref{aug1_6}), $\int_{-\infty }^{\infty }dt\ \Lambda \left(
t, t\right) =I_{m}$. Furthermore, it can be shown that the two-time
correlation $\langle \Psi _{\mathrm{in}}|\breve{b}_{\mathrm{in}}(t)\breve{b}_{\mathrm{in}}^{\dag
}(r)|\Psi _{\mathrm{in}}\rangle$ has the form 
\begin{equation}
\left\langle \Psi _{\mathrm{in}}|\breve{b}_{\mathrm{in}}(t)\breve{b}_{\mathrm{in}}^{\dag
}(r)|\Psi _{\mathrm{in}}\right\rangle =\delta (t-r)\left[ 
\begin{array}{cc}
I_{m} & 0 \\ 
0 & 0_{m}
\end{array}
\right] +\left[ 
\begin{array}{cc}
\Lambda \left( r,t\right) & 0 \\ 
0 & \Lambda \left( t,r\right)
\end{array}
\right].  \label{eq:input_non_Markovian}
\end{equation}

\begin{remark}
If all the input fields are in the vacuum state, i.e., $|\Psi_{\rm in}\rangle = |0^{\otimes m}\rangle$, it is well-known that 
\begin{equation}
\left\langle 0^{\otimes m}|\breve{b}_{\mathrm{in}}(t)\breve{b}_{\mathrm{in}
}^{\dag }(r)|0^{\otimes m}\right\rangle =\delta (t-r)\left[ 
\begin{array}{cc}
I_{m} & 0 \\ 
0 & 0_{m}
\end{array}
\right] .
\end{equation}
In this case, the field is Markovian. The second term on the right-hand side
of Eq. (\ref{eq:input_non_Markovian}) reveals the \textit{non-Markovian}
nature of the $m$-channel $m$-photon input fields. Moreover, due to the
presence of the pulse shape $\psi _{\mathrm{in}}$ in all the diagonal
entries of $\Lambda \left(t, r\right) $, the inputs can be regarded as
correlated non-Markovian noise inputs.
\end{remark}


\subsection{The passive case: output intensity}\label{subsec:intensity}

In this subsection, for the \textit{passive }quantum linear system (\ref
{sys_passive}) driven by an $m$-photon input state $|\Psi _{\mathrm{in}}\rangle $
defined in Eq. (\ref{eq:aug23_10}), we derive a formula for the output
intensity $\bar{n}_{\mathrm{out}}(t)$ defined in Eq. (\ref{n_out2}).

Recall that in the passive case the matrix $C_{+}=0$. Substitution of $
L(t)=C_{-}a(t)$ into Eq. (\ref{eq:output_gauge}) yields 
\begin{equation}
d\Lambda _{\mathrm{out}}(t)=S_{-}^{\#}d\Lambda _{\mathrm{in}
}(t)S_{-}^{T}+S_{-}^{\#}dB_{\mathrm{in}}^{\#}(t)a^{T}(t)C_{-}^{T}+C_{-}^{
\#}a^{\#}(t)dB_{\mathrm{in}}^{T}(t)S_{-}^{T}+C_{-}^{\#}a^{
\#}(t)a^{T}(t)C_{-}^{T}dt.  \label{eq:output_gauge_2}
\end{equation}
Inspired by the second term on the right-hand side of Eq. (\ref
{eq:output_gauge_2}), we define an $n$-by-$m$ matrix function $f(t)$ as 
\begin{equation}
f(t)\triangleq \langle \phi \Psi _{\mathrm{in}}|b_{\mathrm{in}
}^{\#}(t)a^{T}(t)|\phi \Psi _{\mathrm{in}}\rangle ^{T}.  \label{eq:m+2}
\end{equation}
Moreover, define an $n\times n$ matrix function $\Sigma (t)$ to be 
\begin{equation}
\Sigma (t)\triangleq \langle \phi \Psi _{\mathrm{in}}|a(t)a^{\dag }(t)|\phi
\Psi _{\mathrm{in}}\rangle ,\ \ \ t\geq t_{0}.  \label{eq:sigma}
\end{equation} 
Clearly, $\Sigma(t) = \Sigma(t)^\dag$.  

The following theorem is the main result of this subsection, which gives an
explicit procedure for computing the output intensity $\bar{n}_{\mathrm{out}}(t)$.

\begin{theorem}\label{thm:intensity} 
For the \textit{passive }quantum linear system (\ref
{sys_passive}) initialized in the vacuum state $|\phi\rangle$ and  driven by the $m$-channel $m$-photon input state $|\Psi _{\mathrm{in}}\rangle $
defined in Eq. (\ref{eq:aug23_10}),
 the matrix function $f(t)$ defined
in Eq. (\ref{eq:m+2}) has the following form 
\begin{equation}
f(t)=-\int_{t_{0}}^{t}e^{A(t-r)}C_{-}^{\dagger } S_- \Lambda \left(
t, r\right) dr, 
 \label{eq:m_+_dot}
\end{equation}
where the matrix function $ \Lambda (t, r)$ is given in Eq. (\ref{eq:input_Lambda}). The output intensity $\bar{n}_{\mathrm{out}}(t)$ is given by
\begin{equation}  \label{eq:intensity}
\bar{n}_{\mathrm{out}}(t) = S_{-}^{\#}\Lambda \left( t, t\right) S_{-}^{T}
+S_{-}^{\#}f(t)^{T}C_{-}^{T}+C_{-}^{\#}f(t)^{\#}S_{-}^{T}-C_{-}^{\#}C_{-}^{T}+C_{-}^{\#}\Sigma (t)^{T}C_{-}^{T}, 
\end{equation}
in which the covariance function $\Sigma (t)$ solves the following matrix
equation 
\begin{equation}
\dot{\Sigma}(t)=A\Sigma (t)+\Sigma (t)A^{\dagger }+C_{-}^{\dagger
}C_{-}-C_{-}^{\dagger }S_{-}f(t)^{\dagger }-f(t)S_{-}^{\dagger }C_{-}
\label{eq:Sigma_dot_2}
\end{equation}
with the initial condition $\Sigma (t_{0})=I_{n}$.
\end{theorem}


\textbf{Proof.} ~ We prove this theorem in three steps.

\textit{Step 1.} We establish Eq. (\ref{eq:m_+_dot}). Firstly, it can be
readily shown that
\begin{equation}
b_{\mathrm{in}}(t)|\Psi _{\mathrm{in}}\rangle =\left[ 
\begin{array}{c}
|\zeta _{1}(t)\rangle \\ 
\vdots \\ 
|\zeta _{m}(t)\rangle
\end{array}
\right] ,  \label{eq:mar27_1}
\end{equation}
where the following {\it notation}
\begin{equation}
|\zeta _{j}(t)\rangle \equiv \int_{-\infty }^{\infty }\cdots \int_{-\infty
}^{\infty }d\tau _{1}\cdots d\tau _{j-1}d\tau _{j+1}\cdots d\tau _{m}\;\psi
_{\mathrm{in}}(\tau _{1},\ldots ,\tau _{j-1},t,\tau _{j+1},\ldots ,\tau
_{m})\tprod\limits_{k=1,k\neq j}^{m}b_{\mathrm{in},k}^{\ast }(\tau
_{k})|0^{\otimes m}\rangle  \label{eq:zeta_0}
\end{equation}
has been used, ($j=1,\ldots ,m)$.  By the notation in Eq. (\ref{eq:zeta_k}), Eq. (\ref{eq:zeta_0}) can be re-written as
\begin{equation}
|\zeta _{j}(t)\rangle = \int_{-\infty }^{\infty }\cdots \int_{-\infty
}^{\infty }d\tau _{1}\cdots d\tau _{j-1}d\tau _{j+1}\cdots d\tau _{m}\;\zeta _{j}(\tau, t)\tprod\limits_{k=1,k\neq j}^{m}b_{\mathrm{in},k}^{\ast }(\tau
_{k})|0^{\otimes m}\rangle.
  \label{eq:zeta}
\end{equation}
 As a result, 
\begin{equation}
\langle \phi \Psi _{\mathrm{in}}|b_{\mathrm{in}}^{\#}(t)b_{\mathrm{in}
}^{T}(r)|\phi \Psi _{\mathrm{in}}\rangle 
=
\langle \Psi _{\mathrm{in}}|b_{
\mathrm{in}}^{\#}(t)b_{\mathrm{in}}^{T}(r)|\Psi _{\mathrm{in}}\rangle 
=
\left[
\begin{array}{ccc}
\langle \zeta _{1}(t)|\zeta _{1}(r)\rangle &  &  \\ 
& \ddots &  \\ 
&  & \langle \zeta _{m}(t)|\zeta _{m}(r)\rangle
\end{array}
\right] =\Lambda \left( t,r\right) ,  
\label{eq:mar27_2}
\end{equation}
where Eq.  (\ref{eq:input_Lambda}) has been used in the last step. 
Moreover, 
\begin{equation}
\langle \phi \Psi _{\mathrm{in}}|b_{\mathrm{in}}^{\#}(t)a^{T}(t)|\phi \Psi _{
\mathrm{in}}\rangle =\left[ 
\begin{array}{c}
\langle \phi \Psi _{\mathrm{in}}|b_{\mathrm{in},m}^{\ast }(t)a^{T}(t)|\phi
\Psi _{\mathrm{in}}\rangle \\ 
\vdots \\ 
\langle \phi \Psi _{\mathrm{in}}|b_{\mathrm{in},m}^{\ast }(t)a^{T}(t)|\phi
\Psi _{\mathrm{in}}\rangle
\end{array}
\right] =\left[ 
\begin{array}{c}
\langle \phi \zeta _{1}(t)|a^{T}(t)|\phi \Psi _{\mathrm{in}}\rangle \\ 
\vdots \\ 
\langle \phi \zeta _{m}(t)|a^{T}(t)|\phi \Psi _{\mathrm{in}}\rangle
\end{array}
\right] .  \label{eq:mar28_temp2}
\end{equation}
Substituting Eq. (\ref{eq:mar28_temp2}) into Eq. (\ref{eq:m+2}) yields
\begin{equation}
f(t)=\left[ 
\begin{array}{ccc}
\langle \phi \zeta _{1}(t)|a(t)|\phi \Psi _{\mathrm{in}}\rangle & \cdots & 
\langle \phi \zeta _{m}(t)|a(t)|\phi \Psi _{\mathrm{in}}\rangle
\end{array}
\right] .  
\label{eq:mar28_temp3}
\end{equation}
Secondly, solving Eq. (\ref{sys_passive}) we get 
\begin{equation}
a(t)=e^{A(t-t_{0})}a-\int_{t_{0}}^{t}e^{A(t-r)}C_{-}^{\dagger }S_- b_{
\mathrm{in}}(r)dr,\ \ t\geq t_{0}.  
\label{eq:a_t}
\end{equation}
Partition the $n$-by-$m$ matrix function $e^{At}C_{-}^{\dagger }S_-$ into $m$
columns, specifically, 
\begin{equation}
e^{At}C_{-}^{\dagger }S_-=[
\begin{array}{ccc}
c_{1}(t) & \cdots & c_{m}(t)
\end{array}
].  \label{eq:c}
\end{equation}
By Eqs. (\ref{eq:a_t}), (\ref{eq:c}), (\ref{eq:mar27_1}), and (\ref{eq:mar27_2}), we have%
\begin{eqnarray}
&&
\left[ 
\begin{array}{ccc}
\langle \phi \zeta _{1}(t)|a(t)|\phi \Psi _{\mathrm{in}}\rangle & \cdots & 
\langle \phi \zeta _{m}(t)|a(t)|\phi \Psi _{\mathrm{in}}\rangle
\end{array}
\right]  \nonumber \\
&=&
-\left[ 
\begin{array}{ccc}
\langle  \zeta _{1}(t)|\int_{t_{0}}^{t}dr\ e^{A(t-r)}C_{-}^{\dagger }S_-b_{
\mathrm{in}}(r)| \Psi _{\mathrm{in}}\rangle & \cdots & \langle 
\zeta _{m}(t)|\int_{t_{0}}^{t}dr\ e^{A(t-r)}C_{-}^{\dagger }S_-b_{\mathrm{in}
}(r)| \Psi _{\mathrm{in}}\rangle
\end{array}
\right]  \nonumber \\
&=&
-\int_{t_{0}}^{t}\left[ 
\begin{array}{ccc}
\sum_{j=1}^{m}\langle  \zeta _{1}(t)|c_{j}(t-r)b_{\mathrm{in},j}(r)|
\Psi _{\mathrm{in}}\rangle & \cdots & \sum_{j=1}^{m}\langle \zeta
_{m}(t)|c_{j}(t-r)b_{\mathrm{in},j}(r)| \Psi _{\mathrm{in}}\rangle%
\end{array}
\right] dr  \nonumber \\
&=&
-\int_{t_{0}}^{t}\left[ 
\begin{array}{ccc}
\sum_{j=1}^{m}\langle \zeta _{1}(t)|c_{j}(t-r))|\zeta _{j}(r)\rangle & \cdots & \sum_{j=1}^{m}\langle \zeta
_{m}(t)|c_{j}(t-r)|\zeta _{j}(r)\rangle
\end{array}
\right] dr  \nonumber \\
&=&-\int_{t_{0}}^{t}\left[ 
\begin{array}{ccc}
\langle \zeta _{1}(t)|c_{1}(t-r)|\zeta _{1}(r)\rangle & \cdots & \langle
\zeta _{m}(t)|c_{m}(t-r)|\zeta _{m}(r)\rangle
\end{array}
\right] dr  \nonumber \\
&=&-\int_{t_{0}}^{t}\left[ 
\begin{array}{ccc}
c_{1}(t-r) & \cdots & c_{m}(t-r)
\end{array}
\right] \left[ 
\begin{array}{ccc}
\langle \zeta _{1}(t)|\zeta _{1}(r)\rangle &  &  \\ 
& \ddots &  \\ 
&  & \langle \zeta _{m}(t)|\zeta _{m}(r)\rangle
\end{array}
\right] dr  \nonumber \\
&=&-\int_{t_{0}}^{t}e^{A(t-r)}C_{-}^{\dagger }S_-\Lambda \left( t,r\right)
dr.  \label{eq:mar28_temp1}
\end{eqnarray}
Substituting Eq. (\ref{eq:mar28_temp1}) into Eq. (\ref{eq:mar28_temp3})
gives Eq. (\ref{eq:m_+_dot}).

\textit{Step 2.} We establish Eq. (\ref{eq:Sigma_dot_2}). By  It\^{o}
calculus and Eq. (\ref{eq:m+2}), we have 
\begin{eqnarray}
&&
d\Sigma (t)  \nonumber \\
&=&
d\langle \phi \Psi _{\mathrm{in}}|a(t)a^{\dag }(t)|\phi \Psi _{\mathrm{in}
}\rangle  \nonumber \\
&=&
\langle \phi \Psi _{\mathrm{in}}|(da(t))a^{\dag }(t)|\phi \Psi _{\mathrm{
in}}\rangle +\langle \phi \Psi _{\mathrm{in}}|a(t)(da^{\dag }(t))|\phi \Psi
_{\mathrm{in}}\rangle +\langle \phi \Psi _{\mathrm{in}}|(da(t))(da^{\dag
}(t))|\phi \Psi _{\mathrm{in}}\rangle  \nonumber \\
&=&
A\Sigma (t)dt+\Sigma (t)A^{\dagger }dt
-C_{-}^{\dagger }S_{-}\langle \phi \Psi _{\mathrm{in}}|dB_{\mathrm{in}
}(t)a^{\dag }(t)|\phi \Psi _{\mathrm{in}}\rangle -\langle \phi \Psi _{
\mathrm{in}}|a(t)dB_{\mathrm{in}}^{\dagger }(t)|\phi \Psi _{\mathrm{in}
}\rangle S_{-}^{\dagger }C_{-}+C_{-}^{\dagger }C_{-}dt  \nonumber \\
&=&A\Sigma (t)dt+\Sigma (t)A^{\dagger }dt+C_{-}^{\dagger }C_{-}dt  \nonumber \\
&&-C_{-}^{\dagger }S_{-}\left[ 
\begin{array}{c}
\langle \phi \Psi _{\mathrm{in}}|a^{\dag }(t)dB_{\mathrm{in},1}(t)|\phi \Psi
_{\mathrm{in}}\rangle \\ 
\vdots \\ 
\langle \phi \Psi _{\mathrm{in}}|a^{\dag }(t)dB_{\mathrm{in},m}(t)|\phi \Psi
_{\mathrm{in}}\rangle
\end{array}
\right] -\left[ 
\begin{array}{ccc}
\langle \phi \Psi _{\mathrm{in}}|dB_{\mathrm{in},1}^{\ast }(t)a(t)|\phi \Psi
_{\mathrm{in}}\rangle & \cdots & \langle \phi \Psi _{\mathrm{in}}|dB_{
\mathrm{in},m}^{\ast }(t)a(t)|\phi \Psi _{\mathrm{in}}\rangle
\end{array}
\right] S_{-}^{\dagger }C_{-}  \nonumber \\
&=&A\Sigma (t)dt+\Sigma (t)A^{\dagger }dt+C_{-}^{\dagger }C_{-}dt
-C_{-}^{\dagger }S_{-}\left[ 
\begin{array}{ccc}
\langle \phi \zeta _{1}(t)|a(t)|\phi \Psi _{\mathrm{in}}\rangle & \cdots & 
\langle \phi \zeta _{m}(t)|a(t)|\phi \Psi _{\mathrm{in}}\rangle
\end{array}
\right] ^{\dagger }dt  \nonumber \\
&&-\left[ 
\begin{array}{ccc}
\langle \phi \zeta _{1}(t)|a(t)|\phi \Psi _{\mathrm{in}}\rangle & \cdots & 
\langle \phi \zeta _{m}(t)|a(t)|\phi \Psi _{\mathrm{in}}\rangle
\end{array}
\right] S_{-}^{\dagger }C_{-}dt  \nonumber \\
&=&A\Sigma (t)dt+\Sigma (t)A^{\dagger }dt+C_{-}^{\dagger }C_{-}dt
-C_{-}^{\dagger }S_{-}f(t)^{\dagger }dt-f(t)S_{-}^{\dagger }C_{-}dt .
\label{eq:Sigma_dot}
\end{eqnarray}
In Eq. (\ref{eq:Sigma_dot}), the commutation relations $[a_{j}(t),dB_{\mathrm{in}
,k}(t)]=[a_{j}^{\ast }(t),dB_{\mathrm{in},k}(t)]=[a_{j}(t),dB_{\mathrm{in}
,k}^{\ast }(t)]=[a_{j}^{\ast }(t),dB_{\mathrm{in},k}^{\ast }(t)]=0$ ($
j=1\ldots ,n$, $k=1,\ldots ,m$) have been used to get the 4th step, and Eq. (\ref{eq:mar28_temp3}) has been used to get the 6th step (which is the last
step). Dividing both sides of Eq. (\ref{eq:Sigma_dot}) by $dt$ yields Eq. (\ref{eq:Sigma_dot_2}).

\textit{Step 3.} We establish Eq. (\ref{eq:intensity}). By the canonical
commutation relation $[a_{j},a_{k}^{\ast }]=\delta _{jk}$ ($j, k=1,\ldots ,n$
), we have 
\begin{eqnarray}
\Sigma (t) &=&\langle \phi \Psi _{\mathrm{in}}|a(t)a^{\dag }(t)|\phi \Psi _{
\mathrm{in}}\rangle  \nonumber \\
&=&\langle \phi \Psi _{\mathrm{in}}|\left( I+(a^{\#}(t)a^{T}(t))^{T}\right)
|\phi \Psi _{\mathrm{in}}\rangle  \nonumber \\
&=&I+\langle \phi \Psi _{\mathrm{in}}|a^{\#}(t)a^{T}(t)|\phi \Psi _{\mathrm{
in}}\rangle ^{T}.  \label{aug1_temp1}
\end{eqnarray}
This, together with Eqs. (\ref{eq:output_gauge_2}) and (\ref{eq:mar28_temp3}), yields
\begin{eqnarray}
&&\langle \phi \Psi _{\mathrm{in}}|d\Lambda _{\mathrm{out}}(t)|\phi \Psi _{
\mathrm{in}}\rangle  \nonumber \\
&=&S_{-}^{\#}\langle \phi \Psi _{\mathrm{in}}|d\Lambda _{\mathrm{in}
}(t)|\phi \Psi _{\mathrm{in}}\rangle S_{-}^{T}+S_{-}^{\#}\langle \phi \Psi _{
\mathrm{in}}|dB_{\mathrm{in}}^{\#}(t)a^{T}(t)|\phi \Psi _{\mathrm{in}
}\rangle C_{-}^{T}  \nonumber \\
&&+C_{-}^{\#}\langle \phi \Psi _{\mathrm{in}}|a^{\#}(t)dB_{\mathrm{in}
}^{T}(t)|\phi \Psi _{\mathrm{in}}\rangle S_{-}^{T}+C_{-}^{\#}\langle \phi
\Psi _{\mathrm{in}}|a^{\#}(t)a^{T}(t)|\phi \Psi _{\mathrm{in}}\rangle
C_{-}^{T}dt  \nonumber \\
&=&S_{-}^{\#}\langle \Psi _{\mathrm{in}}|d\Lambda _{\mathrm{in}}(t)|\Psi _{
\mathrm{in}}\rangle
S_{-}^{T}+S_{-}^{\#}f(t)^{T}C_{-}^{T}dt+C_{-}^{\#}f(t)^{
\#}S_{-}^{T}dt-C_{-}^{\#}C_{-}^{T}dt+C_{-}^{\#}\Sigma (t)^{T}C_{-}^{T}dt.
\label{eq:temp2bb}
\end{eqnarray}
By Eq. (\ref{eq:mar27_2}),
\begin{equation}
\langle \phi \Psi _{\mathrm{in}}|d\Lambda _{\mathrm{in}}(t)|\phi \Psi _{
\mathrm{in}}\rangle =\Lambda \left( t, t\right) dt.  \label{eq:temp1bb}
\end{equation}
Substituting Eq. (\ref{eq:temp1bb}) into Eq. (\ref{eq:temp2bb}) and dividing
both sides of the resulting equation by $dt$ yield Eq. (\ref{eq:intensity}). $\blacksquare $



\subsection{The passive case: state transfer}\label{sec:passive_mm}

In this subsection, we derive an analytical form of the output field state of the 
\textit{passive} quantum linear system (\ref
{sys_passive}) driven by the $m$-photon input state $|\Psi
_{\mathrm{in}}\rangle $ defined in Eq. (\ref{eq:aug23_10}).

The following is the main result of this subsection.

\begin{theorem} \label{thm:passive_mm}
If the asymptotically stable \textit{passive} quantum linear system (\ref
{sys_passive}) is initialized in the vacuum state and is driven by the $m$-channel $m$-photon input state $|\Psi _{\mathrm{in}}\rangle $
defined in Eq. (\ref{eq:aug23_10}), then the steady-state output field state is an $m$-channel $m$-photon state of the form 
\begin{equation}
|\Psi _{\mathrm{out}}\rangle =\psi _{\mathrm{out}}\circledcirc ^{m}b_{
\mathrm{in}}^{\#}|0^{\otimes m}\rangle , 
 \label{dec12_thm2}
\end{equation}
where the operation $\circledcirc ^{m}$ has been defined in Eq. (\ref{dec3_circ^2}), and the output pulse $\psi _{\mathrm{out}}$ is given by the $m$-fold
convolution
\begin{equation}
\psi _{\mathrm{out},j_{1}\ldots j_{m}}(r_{1},\ldots ,r_{m})=\int_{-\infty
}^{\infty }\cdots \int_{-\infty }^{\infty }dt_{1}\cdots
dt_{m}\;g_{G^{-}}^{j_{1}1}(r_{1}-t_{1})\cdots
g_{G^{-}}^{j_{m}m}(r_{m}-t_{m})\psi _{\mathrm{in}}(t_{1},\ldots ,t_{m}) 
\label{dec23_2}
\end{equation}
($j_{1}, \ldots, j_{m}=1,\ldots, m$) with the impulse response function $g_{G^{-}}(t)$ given in Eq. (\ref{july17_1}). If
we update the $m$-variable function $\psi _{\mathrm{in}}$ in Eq. (\ref{eq:aug23_10}) to a tensor $\psi _{\mathrm{in}}^{\uparrow }$ with entries
\begin{equation}
\psi _{\mathrm{in,}j_{1}\ldots j_{m}}^{\uparrow }(r_{1},\ldots
,r_{m})\triangleq \left\{ 
\begin{array}{ll}
\psi _{\mathrm{in}}(r_{1},\ldots ,r_{m}), & \mathrm{if\ }j_{1}=1,j_{2}=2,
\ldots ,j_{m}=m, \\ 
0, & \mathrm{otherwise},
\end{array}
\right.  \label{july31_5}
\end{equation}%
as has been done in Eq. (\ref{dec_up1}), then the output pulse $\psi _{\mathrm{
out}}$ can be written in a compact form 
\begin{equation}
\psi _{\mathrm{out}}=\psi _{\mathrm{in}}^{\uparrow }\circledast
_{t}^{m}g_{G^{-}},
 \label{eq:aug26_10}
\end{equation}
where the operation $\circledast_{t}^{m}$ has been defined in Eq. (\ref{eq:product_m_b}).
\end{theorem}

\textbf{Proof.}~ To prove this result, we use both the Schr\"{o}dinger picture and  Heisenberg picture.  We first work in  the Heisenberg picture. By Eqs. (\ref{sys_passive}) and (\ref{eq:a_t}),
\begin{equation*}
b_{\mathrm{out}}(t)
=
Ce^{A(t-t_{0})}a-\int_{t_{0}}^{t}Ce^{A(t-r)}C^{\dagger
}S_{-}b_{\mathrm{in}}(r)dr+S_{-}b_{\mathrm{in}}(t)
\end{equation*}
In terms of Eq. (\ref{july17_1}), the above equation can be re-written as
\begin{equation*}
b_{\mathrm{out}}(t)=Ce^{A(t-t_{0})}a+\int_{t_{0}}^{t}g_{G^{-}}(t-r)b_{
\mathrm{in}}(r)dr,\ \ \ t\geq t_{0},
\end{equation*}
whose adjoint operator $b_{\mathrm{out}}^{\#}(t)$ satisfies 
\begin{equation}
b_{\mathrm{out}}^{\#}(t)=C^{\#}e^{A^{\#}(t-t_{0})}a^{\#}+
\int_{t_{0}}^{t}g_{G^{-}}(t-r)^{\#}b_{\mathrm{in}}^{\#}(r)dr,\ \ \ t\geq
t_{0}.  \label{dec11_5}
\end{equation}
On the other hand, notice that in the Heisenberg picture, Eq. (\ref{eq:b_out_dec22}) gives 
\begin{equation}
b_{\mathrm{out}}^{\#}(t)=U(t, t_{0})^{\ast }b_{\mathrm{in}}^{
\#}(t)U(t, t_{0}),\ \ \ t\geq t_{0}  \label{dec11_6}
\end{equation}
(component-wise for the components of $b^\#_{\rm in}(t)$).
Eqs. (\ref{dec11_5})-(\ref{dec11_6}) yield 
\begin{eqnarray}
b_{\mathrm{in}}^{\#}(t)
&=&C^{\#}e^{A^{\#}(t-t_{0})}U(t, t_{0})a^{\#}U(t, t_{0})^{\ast
}+\int_{t_{0}}^{t}g_{G^{-}}(t-r)^{\#}U(t, t_{0})b_{\mathrm{in}
}^{\#}(r)U(t, t_{0})^{\ast }dr  
\nonumber 
\\
&=&C^{\#}e^{A^{\#}(t-t_{0})}U(t, t_{0})a^{\#}U(t, t_{0})^{\ast
}+\int_{t_{0}}^{t}g_{G^{-}}(t-r)^{\#}U(r, t_{0})b_{\mathrm{in}
}^{\#}(r)U(r, t_{0})^{\ast }dr,  
\label{dec11_7}
\end{eqnarray}
where the fact 
\begin{equation*}
U(t, t_{0})b_{\mathrm{in}}^{\#}(r)U(t, t_{0})^{\ast }=U(r, t_{0})b_{\mathrm{in}
}^{\#}(r)U(r, t_{0})^{\ast },\ \ \ t\geq r\geq t_{0}
\end{equation*}
has been used to get the last term on the right-hand side of Eq. (\ref
{dec11_7}). Since the system is asymptotically stable, sending $
t_{0}\rightarrow -\infty $, Eq. (\ref{dec11_7}) becomes 
\begin{equation}
b_{\mathrm{in}}^{\#}(t)=\int_{-\infty }^{t}g_{G^{-}}(t-r)^{\#}U(r,-\infty
)b_{\mathrm{in}}^{\#}(r)U(r,-\infty )^{\ast }dr=\int_{-\infty }^{\infty
}g_{G^{-}}(t-r)^{\#}U(r,-\infty )b_{\mathrm{in}}^{\#}(r)U(r,-\infty )^{\ast
}dr.  \label{dec11_8}
\end{equation}
This, together with Eq. (\ref{dec1_g_inv}), yields 
\begin{equation}
U(t,-\infty )b_{\mathrm{in}}^{\#}(t)U(t,-\infty )^{\ast }
=
\int_{-\infty}^{\infty }g_{G^{-}}(r-t)^{T}b_{\mathrm{in}}^{\#}(r)dr.  
\label{dec11_9}
\end{equation}

Next, we switch to the Schr\"{o}dinger picture.  In the Schr\"{o}dinger picture, the joint system-field state at time $t\geq t_0$ is $U(t, t_0)|\phi \Psi_{\rm in}\rangle$. Thus,  the steady-state output field state can be
obtained by tracing out the system. That is, 
\begin{equation}
|\Psi _{\mathrm{out}}\rangle =\langle \phi |\lim_{t_0\to -\infty, t\to \infty}U(t, t_{0})|\phi \Psi
_{\mathrm{in}}\rangle .  \label{dec13_1}
\end{equation}
We have
\begin{eqnarray}
&&
|\Psi _{\mathrm{out}}\rangle  \label{Dec11_10} 
\\
&=&
\lim_{t_{0}\rightarrow -\infty ,t\rightarrow \infty }\left\langle \phi
\right\vert U(t, t_{0})|\phi \Psi _{\mathrm{in}}\rangle  
\nonumber \\
&=&
\lim_{t_{0}\rightarrow -\infty ,t\rightarrow \infty }\int_{-\infty
}^{\infty }\cdots \int_{-\infty }^{\infty }dt_{1}\cdots dt_{m\ }\psi _{
\mathrm{in}}(t_{1},\ldots ,t_{m})\left\langle \phi \right\vert U(t, t_{0})b_{
\mathrm{in,1}}^{\ast }(t_{1})\cdots b_{\mathrm{in},m}^{\ast }(t_{m})|\phi
0^{\otimes m}\rangle  
\nonumber \\
&=&
\lim_{t_{0}\rightarrow -\infty ,t\rightarrow \infty
}\int_{t_{0}}^{t}\cdots \int_{t_{0}}^{t}dt_{1}\cdots dt_{m\ }\psi _{\mathrm{
in}}(t_{1},\ldots ,t_{m})\left\langle \phi \right\vert U(t, t_{0})b_{\mathrm{
in,1}}^{\ast }(t_{1})\cdots b_{\mathrm{in},m}^{\ast }(t_{m})|\phi 0^{\otimes
m}\rangle  
\nonumber \\
&=&
\lim_{t_{0}\rightarrow -\infty ,t\rightarrow \infty
}\int_{t_{0}}^{t}\cdots \int_{t_{0}}^{t}dt_{1}\cdots dt_{m}\ \psi _{\mathrm{
in}}(t_{1},\ldots ,t_{m})\left\langle \phi \right\vert U(t_{1}, t_{0})b_{
\mathrm{in,1}}^{\ast }(t_{1})U(t_{1},t_{0})^{\ast }  
\nonumber \\
&&
\ \ \cdots U(t_{m},t_{0})b_{\mathrm{in},m}^{\ast }(t_{m})U(t_{m},t_{0})^{\ast
}U(t_{m},t_{0})|\phi 0^{\otimes m}\rangle  
\nonumber \\
&=&
\lim_{t_{0}\rightarrow -\infty ,t\rightarrow \infty
}\int_{t_{0}}^{t}\cdots \int_{t_{0}}^{t}dt_{1}\cdots dt_{m}\ \psi _{\mathrm{
in}}(t_{1},\ldots ,t_{m})\left\langle \phi \right\vert U(t_{1}, t_{0})b_{
\mathrm{in,1}}^{\ast }(t_{1})U(t_{1},t_{0})^{\ast }  
\nonumber \\
&&
\ \ \cdots U(t_{m},t_{0})b_{\mathrm{in},m}^{\ast }(t_{m})U(t_{m}, t_{0})^{\ast
}|\phi 0^{\otimes m}\rangle  
\label{dec11_11} \\
&=&
\lim_{t\rightarrow \infty }\int_{-\infty }^{t}\cdots \int_{-\infty
}^{t}dt_{1}\cdots dt_{m}\ \psi _{\mathrm{in}}(t_{1},\ldots
,t_{m})\left\langle \phi \right\vert U(t_{1},-\infty )b_{\mathrm{in,1}
}^{\ast }(t_{1})U(t_{1}, -\infty )^{\ast }  
\nonumber \\
&&
\ \ \cdots U(t_{m}, -\infty )b_{\mathrm{in},m}^{\ast }(t_{m})U(t_{m},-\infty
)^{\ast }|\phi 0^{\otimes m}\rangle  
\nonumber \\
&=&
\int_{-\infty }^{\infty }\cdots \int_{-\infty }^{\infty }dt_{1}\cdots
dt_{m}\ \psi _{\mathrm{in}}(t_{1},\ldots ,t_{m})\left\langle \phi
\right\vert \int_{-\infty }^{\infty
}\sum_{j_{1}=1}^{m}g_{G^{-}}^{j_{1}1}(r_{1}-t_{1})b_{\mathrm{in}
,j_{1}}^{\ast }(r_{1})dr_{1} 
 \nonumber \\
&&
\ \ \cdots \int_{-\infty }^{\infty
}\sum_{j_{m}=1}^{m}g_{G^{-}}^{j_{m}m}(r_{m}-t_{m})b_{\mathrm{in}
,j_{m}}^{\ast }(r_{m})dr_{m}|\phi 0^{\otimes m}\rangle  
\label{dec27_1} \\
&=&
\sum_{j_{1}=1}^{m}\cdots \sum_{j_{m}=1}^{m}\int d\overrightarrow{r}\;b_{
\mathrm{in},j_{1}}^{\ast }(r_{1})\cdots b_{\mathrm{in},j_{m}}^{\ast }(r_{m}) \ 
\boxed{\int d\overrightarrow{t}\;g_{G^{-}}^{j_{1}1}(r_{1}-t_{1})\cdots
g_{G^{-}}^{j_{m}m}(r_{m}-t_{m})\psi _{\mathrm{in}}(t_{1},\ldots
,t_{m})} \ |0^{\otimes m}\rangle  
\nonumber \\
&=&
\sum_{j_{1},\ldots ,j_{m}=1}^{m}\int d\overrightarrow{r}\;b_{\mathrm{in}
,j_{1}}^{\ast }(r_{1})\cdots b_{\mathrm{in},j_{m}}^{\ast }(r_{m})\ \boxed{
\psi _{\mathrm{out},j_{1}\ldots j_{m}}(r_{1},\ldots ,r_{m})}\ |0^{\otimes
m}\rangle  
\label{dec11_14}\\
&=&
\psi _{\mathrm{out}}\circledcirc b_{\mathrm{in}}^{\#}\ |0^{\otimes
m}\rangle , 
\nonumber 
\end{eqnarray}
which is exactly Eq. (\ref{dec12_thm2}).  
Notice that the following fact,
\begin{equation}
U(t, t_{0})|\phi 0^{\otimes m}\rangle =|\phi 0^{\otimes m}\rangle ,\ \ \
t\geq t_{0}  \label{dec11_12}
\end{equation}
upon a global phase, has been used to get Eq. (\ref{dec11_11}) from the previous step.  In fact, Eq. (\ref{dec11_12}) holds for general passive
systems, see, e.g., \cite[Lemma 3]{PZJ15}. Eq. (\ref{dec11_9})  has been used to derive Eq. (\ref{dec27_1}). Finally, by the two terms highlighted in the two boxes above, it is clear that 
\[
\psi _{\mathrm{out},j_{1}\ldots j_{m}}(r_{1},\ldots ,r_{m}) = \int d\overrightarrow{t}\;g_{G^{-}}^{j_{1}1}(r_{1}-t_{1})\cdots
g_{G^{-}}^{j_{m}m}(r_{m}-t_{m})\psi _{\mathrm{in}}(t_{1},\ldots
,t_{m}),
\]
which is exactly Eq. (\ref{dec23_2}).  $\blacksquare $

\begin{remark}\label{rem:SH}
In quantum mechanics, the Schr\"{o}dinger picture describes how quantum states evolve; on the other hand, the Heisenberg picture describes how operators evolve. Eq. (\ref{dec12_thm2}) tells us how the input state $|\Psi_{\rm in}\rangle$ evolves and becomes the output state $|\Psi_{\rm out}\rangle$. That is, it is in the Schr\"{o}dinger picture. In the Schr\"{o}dinger picture, operators do not evolve. This is the reason why the input operator $b_{\rm in}^\#$ appears in Eq. (\ref{dec12_thm2}).
\end{remark}

\begin{remark}
When the input pulse is of a product form 
\begin{equation}
\psi _{\mathrm{in}}(t_{1},\ldots ,t_{m})=\xi _{1}(t_{1})\cdots \xi
_{m}(t_{m}),  \label{july30_1}
\end{equation}
the input state $|\Psi _{\mathrm{in}}\rangle $ in Eq. (\ref{eq:aug23_10})
becomes a separable state 
\begin{equation}
|\Psi _{\mathrm{in}}\rangle =\tprod\limits_{k=1}^{m}\mathbf{B}_{\mathrm{in}
,k}^{\ast }(\xi _{k})|0_{k}\rangle ,  \label{july30_2}
\end{equation}
where the notation 
\begin{equation}
\mathbf{B}_{\mathrm{in},k}^{\ast }(\xi ) \equiv \int_{-\infty }^{\infty
}\xi (t)b_{\mathrm{in},k}^{\ast }(t)dt , ~~k=1,\ldots ,m 
\label{july31_B}
\end{equation}
has been used. In this case, by Eq. (\ref{dec23_2})
we have
\begin{equation}
\psi _{\mathrm{out},j_{1}\ldots j_{m}}(r_{1},\ldots
,r_{m})=\tprod\limits_{k=1}^{m}\int_{-\infty }^{\infty }
g_{G^{-}}^{j_{k}k}(r_{k}-t_{k})\xi _{k}(t_{k}) dt_{k}, ~~ j_{1},\ldots
,j_{m}=1,\ldots ,m.  \label{july30_3}
\end{equation}
Define 
\begin{equation}
\xi _{\mathrm{out},jk}(r)\triangleq \int_{-\infty }^{\infty
}g_{G^{-}}^{jk}(r-t)\xi _{k}(t)dt,\ \ j,k=1,\ldots ,m.  \label{july30_4}
\end{equation}
Then, by Theorem \ref{thm:passive_mm} and Eq. (\ref{july30_3}),
\begin{equation}
|\Psi _{\mathrm{out}}\rangle =\sum_{j_{1},\ldots
,j_{m}=1}^{m}\tprod\limits_{k=1}^{m}\mathbf{B}_{\mathrm{in},j_{k}}^{\ast
}(\xi _{\mathrm{out},j_k k})|0^{\otimes m}\rangle
=\tprod\limits_{k=1}^{m}\sum_{j=1}^{m}\mathbf{B}_{\mathrm{in},j}^{\ast
}(\xi _{\mathrm{out},jk})|0^{\otimes m}\rangle . 
 \label{july30_5}
\end{equation}
 Interestingly, $|\Psi _{\mathrm{out}}\rangle $ in Eq. (\ref
{july30_5}) can also be derived by means of \cite[Theorem 5]{zhang13}.
Therefore, Theorem \ref{thm:passive_mm} generalizes one of the main results
in \cite{zhang13}.
\end{remark}

\begin{figure}[ptb]
\centering
\includegraphics[width=2.5in]{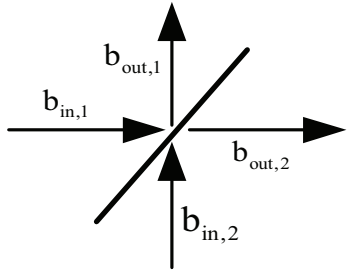}
\caption{Schematic representation of a beamsplitter}
\label{fig:BS}
\end{figure}

\begin{example}
\label{ex:BS0} (beamsplitter.) A beamsplitter is a static device
widely used in optical laboratories, \cite{Leo03}, \cite{BR04}, \cite{NJD09}%
, see Fig.~\ref{fig:BS}. In the $(S_{-},L,H)$ formalism, a
beamsplitter may be modeled by $L=0$, $H=0$, and 
\begin{equation}
S_{-}=\left[ 
\begin{array}{cc}
R & T \\ 
T & R
\end{array}
\right] ,~~R,T\in \mathbb{C},~|R|^{2}+|T|^{2}=1.  
\label{july31_bs}
\end{equation}
Let the 2-channel 2-photon input state be 
\begin{equation}
|\Psi _{\mathrm{in}}\rangle =\int_{-\infty }^{\infty }\int_{-\infty
}^{\infty }dt_{1}dt_{2}\;\psi _{\mathrm{in}}(t_{1},t_{2})b_{\mathrm{in},1}^{\ast }(t_{1})b_{\mathrm{in},2}^{\ast }(t_{2})|0_{1}0_{2}\rangle .
\label{eq:aug16_2}
\end{equation}
By Theorem \ref{thm:passive_mm}, the steady-state output field state is
\begin{eqnarray}
&&|\Psi _{\mathrm{out}}\rangle  \nonumber \\
&=&RT\int dr_{1}dr_{2}\ b_{\mathrm{in},1}^{\ast }(r_{1})b_{\mathrm{in},1}^{\ast }(r_{2})\psi _{\mathrm{in}}(r_{1},r_{2})|0_{1}\rangle \otimes
|0_{2}\rangle +R^{2}\int dr_{1}dr_{2}\ b_{\mathrm{in},1}^{\ast }(r_{1})b_{\mathrm{in},2}^{\ast }(r_{2})\psi _{\mathrm{in}}(r_{1},r_{2})|0_{1}\rangle
\otimes |0_{2}\rangle  \nonumber \\
&&+T^{2}\int dr_{1}dr_{2}\ b_{\mathrm{in},1}^{\ast }(r_{1})b_{\mathrm{in},2}^{\ast }(r_{2})\psi _{\mathrm{in}}(r_{2},r_{1})|0_{1}\rangle \otimes
|0_{2}\rangle +|0_{1}\rangle \otimes RT\int dr_{1}dr_{2}\ b_{\mathrm{in},2}^{\ast }(r_{1})b_{\mathrm{in},2}^{\ast }(r_{2})\psi _{\mathrm{in}}(r_{1},r_{2})|0_{2}\rangle , 
\nonumber
\end{eqnarray}
which is exactly \cite[Eq. (6.8.7)]{RL00}.
\end{example}


\begin{example}
\label{ex:entangled} (optical cavity.) An optical cavity is a system
composed of reflecting and/or transmitting mirrors \cite[Chapter 5.3]{BR04}, 
\cite[Chapter 7]{WM08}, \cite{GJ09}, \cite{NJD09}. A widely used type of
optical cavities is the so-called Fabry-Perot cavity. In the $(S_{-},L,H)$
formalism, a single-mode Fabry-Perot cavity with two input channels, as
shown in Fig. \ref{fig:cavity}, can be modeled with parameters 
\begin{equation}
\left( S_- =I_2, ~ L=\left[ 
\begin{array}{c}
\sqrt{\kappa _{1}}a \\ 
\sqrt{\kappa _{2}}a
\end{array}
\right] , ~H=\omega _{d}a^{\ast }a\right) .  
\label{eq:aug23_cavity}
\end{equation}
Here, $\kappa _{1}$ and $\kappa _{2}$ are coupling strengths between the
cavity and the external fields, and $\omega _{d}$ is the detuning frequency between
the resonant frequency of the cavity and the external fields. (Here we
assume that the two input light fields have the same carrier frequency.) 
\begin{figure}[tbp]
\centering
\includegraphics[width=2.5in]{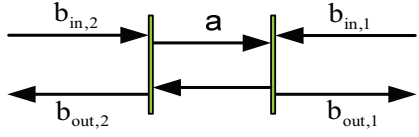}
\caption{Schematic representation of a single-mode Fabry-Perot cavity with
two inputs}
\label{fig:cavity}
\end{figure}
By Eq. (\ref{sys_passive}) we have the following QSDEs 
\begin{eqnarray}
\dot{a}(t) &=&-(\frac{\kappa _{1}+\kappa _{2}}{2}+\mathrm{i}\omega _{d})a(t)-\sqrt{\kappa _{1}}b_{\mathrm{in},1}(t)-\sqrt{\kappa _{2}}b_{\mathrm{in},2}(t),  \nonumber \\
b_{\mathrm{out},1}(t) &=&\sqrt{\kappa _{1}}a(t)+b_{\mathrm{in},1}(t),  \nonumber
\\
b_{\mathrm{out},2}(t) &=&\sqrt{\kappa _{2}}a(t)+b_{\mathrm{in},2}(t).
\label{july31_cavity}
\end{eqnarray}
Let the input state be that given in Eq. (\ref{eq:aug16_2}). In what
follows, we calculate the steady-state output field state. Define the
following two-variable functions 
\begin{equation}
\Phi _{1}(r,t_{2})\triangleq \int_{-\infty }^{r}dt_{1}\ e^{-\left( \mathrm{i}\omega _{d}+\frac{\kappa _{1}+\kappa _{2}}{2}\right) (r-t_{1})}\psi _{\mathrm{in}}(t_{1},t_{2}),  \label{eq:aug13_8}
\end{equation}
\begin{equation}
\Phi _{2}(t_{1},r)\triangleq \int_{-\infty }^{r}dt_{2}\ e^{-\left( \mathrm{i}\omega _{d}+\frac{\kappa _{1}+\kappa _{2}}{2}\right) (r-t_{2})}\psi _{\mathrm{in}}(t_{1},t_{2}),  \label{eq:aug13_7}
\end{equation}
and
\begin{equation}
\Phi (r,\tau )\triangleq \int_{-\infty }^{r}\int_{-\infty }^{\tau
}dt_{1}dt_{2}\ e^{-\left( \mathrm{i}\omega _{d}+\frac{\kappa _{1}+\kappa _{2}}{2}\right) (r+\tau -t_{1}-t_{2})}\psi _{\mathrm{in}}(t_{1},t_{2}).
\label{eq:aug13_10}
\end{equation}
By Theorem \ref{thm:passive_mm}, the steady-state output field state is
\begin{eqnarray}
&&
|\Psi _{\mathrm{out}}\rangle  
\nonumber \\
&=&
\sqrt{\kappa _{1}\kappa _{2}}\int_{-\infty }^{\infty }\int_{-\infty
}^{\infty }dr_{1}dr_{2}\;b_{\mathrm{in},1}^{\ast }(r_{1})b_{\mathrm{in},1}^{\ast }(r_{2})[\kappa _{1}\Phi (r_{1},r_{2})-\Phi
_{2}(r_{1},r_{2})]|0_{1}\rangle \otimes |0_{2}\rangle  
\nonumber \\
&&
+\int_{-\infty }^{\infty }\int_{-\infty }^{\infty }dr_{1}dr_{2}\;b_{\mathrm{in},1}^{\ast }(r_{1})b_{\mathrm{in},2}^{\ast }(r_{2})\left[ \psi _{\mathrm{in}}(r_{1},r_{2})-\kappa _{1}\Phi _{1}(r_{1},r_{2})-\kappa _{2}\Phi
_{2}(r_{1},r_{2})\right.  
\nonumber \\
&&\ \ \ \left. +\kappa _{1}\kappa _{2}(\Phi (r_{1},r_{2})+\Phi
(r_{2},r_{1})) \right] |0_{1}\rangle \otimes |0_{2}\rangle  
\nonumber \\
&&+|0_{1}\rangle \otimes \sqrt{\kappa _{1}\kappa _{2}}\int_{-\infty
}^{\infty }\int_{-\infty }^{\infty }dr_{1}dr_{2}\;b_{\mathrm{in},2}^{\ast
}(r_{1})b_{\mathrm{in},2}^{\ast }(r_{2})\left[ \kappa _{2}\Phi
(r_{1},r_{2})-\Phi _{1}(r_{1},r_{2})\right] |0_{2}\rangle .
\label{eq:aug23_11}
\end{eqnarray}
In what follows we discuss two cases.

Case 1) In the limit $\kappa _{1}\rightarrow 0$, the state in Eq. (\ref{eq:aug23_11}) becomes 
\begin{equation}
|\Psi _{\mathrm{out}}\rangle =\int_{-\infty
}^{\infty }\int_{-\infty }^{\infty } dr_{1}dr_{2}\ b_{\mathrm{in},1}^{\ast
}(r_{1})b_{\mathrm{in},2}^{\ast }(r_{2})[\psi _{\mathrm{in}}(r_{1},r_{2})-\kappa _{2}\Phi _{2}(r_{1},r_{2})]|0_{1}\rangle \otimes
|0_{2}\rangle .  
\label{july31_2}
\end{equation}
If the input field state is an entangled state, the output field state in Eq. (\ref{july31_2}) is also an entangled state. Therefore, even though the system does not affect the first channel directly because of $\kappa_1=0$, it does influence the first channel via its influence on the second channel.

Case 2) The input state is a product state. Assume
\begin{equation}
\psi _{\mathrm{in}}(t_{1},t_{2})=\xi _{1}(t_{1})\xi _{2}(t_{2}),
\label{eq:aug11_5}
\end{equation}
that is, the input is a tensor product state of two single-photon states,
one for each channel. In this case, there exists no entanglement between the two input channels. For this product state, Eqs. (\ref{eq:aug13_8})-(\ref{eq:aug13_10})
reduce to 
\begin{equation}
\Phi _{1}(r_{1},r_{2})=\xi _{2}(r_{2})\eta _{1}(r_{1}),  \label{aug1_9}
\end{equation}
\begin{equation}
\Phi _{2}(r_{1},r_{2})=\xi _{1}(r_{1})\eta _{2}(r_{2}),  \label{aug1_10}
\end{equation}
and
\begin{equation}
\Phi (r_{1},r_{2})=\eta _{1}(r_{1})\eta _{2}(r_{2}),  \label{aug1_11}
\end{equation}
where
\begin{equation}
\eta _{i}(t)\triangleq \int_{-\infty }^{t}e^{-\left( \mathrm{i}\omega _{d}+
\frac{\kappa _{1}+\kappa _{2}}{2}\right) (t-r)}\xi _{i}(r)dr,\ i=1,2.
\label{aug1_12}
\end{equation}
As a result, Eq. (\ref{eq:aug23_11}) becomes 
\begin{equation}
|\Psi _{\mathrm{out}}\rangle 
=
\left( \mathbf{B}_{\mathrm{in,}1}^{\ast
}\left( \xi _{1} - \kappa_1 \eta_1\right)  -\sqrt{\kappa _{1}\kappa _{2}}\mathbf{B}_{\mathrm{in},2}^{\ast }\left( \eta _{2}\right) \right) \left( \mathbf{B}_{\mathrm{in},2}^{\ast }\left( \xi _{2} - \kappa_2 \eta_2\right) -\sqrt{\kappa _{1}\kappa _{2}}\mathbf{B}_{\mathrm{in,}1}^{\ast }\left( \eta _{1}\right)  \right) |0_{1}\rangle \otimes
|0_{2}\rangle   \label{eq:aug13_5}
\end{equation}
 The state in Eq.
(\ref{eq:aug13_5}) is an entangled state. Therefore, the system entangled the initially separable input state. Sending $\kappa _{1}\rightarrow 0$
in Eq. (\ref{eq:aug13_5}) yields 
\begin{equation}
|\Psi _{\mathrm{out}}\rangle =\mathbf{B}_{\mathrm{in,}1}^{\ast }\left( \xi
_{1}\right) |0_{1}\rangle \otimes \mathbf{B}_{\mathrm{in,}2}^{\ast }\left(
\xi _{2}-\kappa _{2}\eta _{2}\right) |0_{2}\rangle ,  
\label{july31_1}
\end{equation}
which is a product state. That is, if the coupling between the system and
the first channel is extremely weak, then the output fields are almost in a
product state. This is reasonable: when the coupling strength $\kappa
_{1}= 0$, the first channel has no interaction with the system, so
the state of the first channel does not change if it is not initially entangled with the second channel. On the other hand, the pulse shape of the second channel has been transformed by the system from $\xi_2$ to $\xi_2-\kappa_2 \eta$.
\end{example}


\subsection{The passive case: the invariant set} \label{subsec:invariant}

Define a class of $m$-channel $m$-photon states of the form 
\begin{eqnarray}
\mathcal{F}_{1} &\triangleq &\Bigg\{|\Psi \rangle =\psi \circledcirc ^{m}b_{\rm in}^{\#}|0^{\otimes m}\rangle \ |\mathrm{the~}m\text{\textrm{-}}\mathrm{way}~m\text{\textrm{-}}\mathrm{dimensional~tensor~function~}\psi \mathrm{~normalizes~}
|\Psi \rangle \Bigg\}.  \label{may1_1}
\end{eqnarray}%

By means of Eq. (\ref{july31_5}), it is clear that the $m$-channel $m$-photon input field state defined in Eq. (\ref
{eq:aug23_10}) can be re-written as $|\Psi_{\rm in}\rangle  = \psi _{\mathrm{in}}^{\uparrow
}  \circledcirc ^{m}b_{\rm in}^{\#}|0^{\otimes m} \rangle$.  Therefore,   $|\Psi_{\rm in}\rangle   \in \mathcal{F}_{1}$.  On the other hand, by Theorem \ref
{thm:passive_mm}, the steady-state output field state $\left\vert \psi _{\mathrm{out}}\right\rangle \in \mathcal{F}_{1}$ too. This motivates us to study more
general pulse shape transfer than that in Theorem \ref{thm:passive_mm}.

The following is the main result of this  subsection.

\begin{theorem}\label{thm:passive_invariance}
 Let the input state for the asymptotically
stable passive quantum linear system (\ref{sys_passive}) (initialized in the vacuum state) be an element $|\Psi _{\mathrm{in}
}\rangle \in \mathcal{F}_{1}$ with pulse shape parametrized by an $m$-way $m$-dimensional tensor function $\psi _{\mathrm{in}}$. Then, the steady-state
output field state
 \begin{equation}
|\Psi _{\mathrm{out}}\rangle =\psi _{\mathrm{out}}\circledcirc ^{m}b_{
\mathrm{in}}^{\#}|0^{\otimes m}\rangle   \label{eq:pulse_transfer_mm}
\end{equation}
 is also an element in $\mathcal{F}_{1}$, where the pulse shape is given by
\begin{equation}
\psi _{\mathrm{out}}=\psi _{\mathrm{in}}\circledast _{t}^{m}g_{G^{-}}.
\label{dec23_3}
\end{equation}
Alternatively, in the frequency domain,
\begin{equation}
\psi _{\mathrm{out}}=\psi _{\mathrm{in}}\circledast _{\omega }^{m}g_{G^{-}}.
\label{eq:pulse_transfer_mm_omega}
\end{equation}
Moreover, we have
\begin{equation}
\left\Vert \psi _{\mathrm{out}}(\mathrm{i}\omega _{1},\ldots ,\mathrm{i}
\omega _{m})\right\Vert ^{2}=\left\Vert \psi _{\mathrm{in}}(\mathrm{i}\omega
_{1},\ldots ,\mathrm{i}\omega _{m})\right\Vert ^{2}, ~~\forall \omega _{1},\ldots ,\omega _{m}\in \mathbb{R}. 
\label{eq:norm}
\end{equation}
\end{theorem}

\textbf{Proof.~} By analogy with the proof of Theorem \ref{thm:passive_mm},
we have 
\begin{eqnarray}
&&
|\Psi _{\mathrm{out}}\rangle  
\nonumber \\
&=&
\lim_{t_{0}\rightarrow -\infty ,t\rightarrow \infty }\left\langle \phi
\right\vert U(t, t_{0})|\phi \Psi _{\mathrm{in}}\rangle  
\nonumber \\
&=&
\lim_{t_{0}\rightarrow -\infty ,t\rightarrow \infty }\sum_{j_{1},\ldots
,j_{m}=1}^{m}\int d\overrightarrow{t}\ \psi _{\mathrm{in},j_{1}\ldots
j_{m}}(t_{1},\ldots ,t_{m})\left\langle \phi \right\vert U(t, t_{0})b_{\mathrm{in},j_{1}}^{\ast }(t_{1})\cdots b_{\mathrm{in},j_{m}}^{\ast
}(t_{m})|\phi 0^{\otimes m}\rangle  
\nonumber \\
&=&
\lim_{t_{0}\rightarrow -\infty ,t\rightarrow \infty }\sum_{j_{1},\ldots
,j_{m}=1}^{m}\int_{t_0}^t \cdots \int_{t_0}^t dt_1 \cdots dt_m  \ \psi _{\mathrm{in},j_{1}\ldots
j_{m}}(t_{1},\ldots ,t_{m})\left\langle \phi \right\vert
U(t_1,t_{0})b_{\mathrm{in},j_{1}}^{\ast}(t_1)U(t_1,t_{0})^{\ast } 
 \nonumber \\
&&
\cdots U(t_m, t_{0})b_{\mathrm{in},j_{m}}^{\ast}(t_m)U(t_m, t_{0})^{\ast }|\phi 0^{\otimes m}\rangle  
\nonumber \\
&=&
\lim_{t\rightarrow \infty }\sum_{j_{1},\ldots
,j_{m}=1}^{m}\int_{-\infty}^t \cdots \int_{-\infty}^t dt_1 \cdots dt_m \  \psi _{\mathrm{in},j_{1}\ldots
j_{m}}(t_{1},\ldots ,t_{m})\left\langle \phi \right\vert U(t_1,-\infty
)b_{\mathrm{in,}j_{1}}^{\ast }(t_1)U(t_1,-\infty )^{\ast } 
 \nonumber \\
&&
\cdots U(t_m,-\infty )b_{\mathrm{in},j_{m}}^{\ast}(t_m)U(t_m,-\infty )^{\ast }|\phi 0^{\otimes m}\rangle  
\nonumber \\
&=&
\sum_{j_{1},\ldots
,j_{m}=1}^{m}\int d\overrightarrow{t}\ \psi _{\mathrm{in},j_{1}\ldots
j_{m}}(t_{1},\ldots ,t_{m})\int_{-\infty }^{\infty
}\sum_{i_{1}=1}^{m}\int_{-\infty }^{\infty
}g_{G^{-}}^{i_{1}j_{1}}(r_{1}-t_{1})b_{\mathrm{in},i_{1}}^{\ast
}(r_{1})dr_{1}  
\nonumber \\
&&
\cdots \int_{-\infty }^{\infty
}\sum_{i_{m}=1}^{m}g_{G^{-}}^{i_{m}j_{m}}(r_{m}-t_{m})b_{\mathrm{in}
,i_{m}}^{\ast }(r_{m})dr_{m}|0^{\otimes m}\rangle  
\nonumber \\
&=&
\sum_{i_{1},\ldots ,i_{m}=1}^{m}\int d\overrightarrow{r}\ b_{\mathrm{in}
,i_{1}}^{\ast }(r_{1})\cdots b_{\mathrm{in},i_{m}}^{\ast
}(r_{m}) 
\nonumber
\\
&&\times \  \boxed{\sum_{j_{1},\ldots ,j_{m}=1}^{m}\int d\overrightarrow{t}\
g_{G^{-}}^{i_{1}j_{1}}(r_{1}-t_{1})\cdots
g_{G^{-}}^{i_{m}j_{m}}(r_{m}-t_{m})\psi _{\mathrm{in},j_{1}\ldots
j_{m}}(t_{1},\ldots ,t_{m})} \left\vert 0^{\otimes
m}\right\rangle
  \nonumber \\
&=&
\sum_{i_{1},\ldots ,i_{m}=1}^{m}\int d\overrightarrow{r}\ \boxed{\psi _{\mathrm{out},i_{1}\ldots i_{m}}(r_{1},\ldots ,r_{m})} \  b_{\mathrm{in},i_{1}}^{\ast
}(r_{1})\cdots b_{\mathrm{in},i_{m}}^{\ast }(r_{m})\left\vert 0^{\otimes
m}\right\rangle ,  \label{may1_3}
\end{eqnarray}
where
\begin{equation}
\psi _{\mathrm{out},i_{1}\ldots i_{m}}(r_{1},\ldots
,r_{m})=\sum_{j_{1},\ldots ,j_{m}=1}^{m}\int d\overrightarrow{t}\
g_{G^{-}}^{i_{1}j_{1}}(r_{1}-t_{1})\cdots
g_{G^{-}}^{i_{m}j_{m}}(r_{m}-t_{m})\psi _{\mathrm{in},j_{1}\ldots
j_{m}}(t_{1},\ldots ,t_{m}),  \label{july18_1}
\end{equation}
as highlighted in the boxes above.  In the compact form, Eq. (\ref{july18_1}) gives Eq. (\ref{dec23_3}).
Therefore, Eq. (\ref{eq:pulse_transfer_mm}) is established. Applying the $m$
-dimensional Fourier transform (\ref{eq:fourier}) to Eq. (\ref
{dec23_3}) gives Eq. (\ref{eq:pulse_transfer_mm_omega}).
Because the system is passive, $G[\mathrm{i}\omega ]$ is a unitary matrix
for all $\omega \in \mathbb{R}$. Consequently, Eq. (\ref{eq:norm}) follows Eq. (\ref{eq:pulse_transfer_mm_omega}) and 
Lemma \ref{lem:norm} immediately. $\blacksquare $

\subsection{The non-passive case}\label{subsec:nonpassive}

A quantum linear system is said to be \textit{non-passive} if $C_{+}\neq 0$
and/or $\Omega _{+}$ $\neq 0$ in Eq. (\ref{eq:ABCD}). Non-passive
elements, such as optical parametric oscillators (OPOs), are key ingredients of
quantum optical systems, \cite{Leo03}, \cite{BR04}, \cite{NJD09}. In this
subsection, we study the output field state of a non-passive quantum linear system driven
by the $m$-channel $m$-photon input field state $|\Psi _{\mathrm{in}}\rangle $ defined in Eq. (\ref{eq:aug23_10}). 

Firstly, we introduce some notation. Define operators
\begin{equation}
b_{j}^{d}(t)\triangleq \left\{ 
\begin{array}{cc}
b_{\mathrm{in,}j}^{\ast }(t), & d=-1, \\ 
b_{\mathrm{in,}j}(t), & d=1,
\end{array}
\right. ,\ \ j=1,\ldots ,m.  \label{eq:aug15_4}
\end{equation}
Then, define an $\underbrace{m\times \cdots \times m}_m \times \underbrace{2\times\cdots \times 2}_m$ tensor
operator $\mathfrak{b}=\mathfrak{b}_{j_{1}\ldots j_{m}}^{d_{1} \ldots
  d_{m}}(t_{1},\ldots ,t_{m})$, whose entries are
\begin{equation}
\mathfrak{b}_{j_{1}\ldots j_{m}}^{d_{1}\ldots d_{m}}(t_{1},\ldots
,t_{m})\triangleq b_{j_{1}}^{d_{1}}(t_{1})\cdots b_{j_{m}}^{d_{m}}(t_{m}),\
\ j_{1},\ldots ,j_{m}=1,\ldots ,m,\ d_{1},\ldots ,d_{m}=\pm 1.
\label{may1_6}
\end{equation}
Denote
\begin{equation}
g_{G^{d}}^{kj}(t)\triangleq \left\{ 
\begin{array}{cc}
g_{G^{-}}^{kj}(t), & d=-1, \\ 
-g_{G^{+}}^{kj}(t)^{\ast }, & d=1,
\end{array}
\right. ,\ \ j,k=1,\ldots ,m.  \label{eq:aug15_5}
\end{equation}
Then, define an  $\underbrace{m\times \cdots \times m}_m \times \underbrace{2\times\cdots \times 2}_m$
tensor function $\psi $ with entries 
\begin{equation}
\psi _{j_{1}\ldots j_{m}}^{d_{1}\ldots d_{m}}(r_{1},\ldots ,r_{m})\triangleq
\int_{-\infty }^{\infty }dt_{1}\cdots dt_{m}\
g_{G^{d_{1}}}^{j_{1}1}(r_{1}-t_{1})\cdots
g_{G^{d_{m}}}^{j_{m}m}(r_{m}-t_{m})\psi _{\mathrm{in}}(t_{1},\ldots ,t_{m})
\label{eq:aug15_6}
\end{equation}
($ j_{1},\ldots ,j_{m}=1,\ldots ,m,\ d_{1},\ldots ,d_{m}=\pm 1$).  
Finally, define the following operation between tensors $\mathfrak{b}$ and $\psi $ 
\begin{equation}
\left\langle \mathfrak{b},\psi \right\rangle \triangleq \sum_{j_{1}, \ldots, j_{m}=1}^{m}\sum_{d_{1}, \ldots, d_{m}=\pm 1}\int dr_{1}\cdots dr_{m} \  \ 
\psi _{j_{1}\ldots j_{m}}^{d_{1}\ldots d_{m}}(r_{1}, \ldots, r_{m})b_{j_{1}}^{d_{1}}(r_{1}) \cdots b_{j_{m}}^{d_{m}}(r_{m}).
\label{eq:inner}
\end{equation}

The following result shows how a non-passive quantum linear system processes
$m$-channel $m$-photon input states.


\begin{theorem}
\label{thm:nonpassive} Let $G$ be an asymptotically stable non-passive  quantum linear
system which is initialized in the vacuum state $|\phi\rangle$ and is driven by the $m$-channel $m$-photon input state $|\Psi _{\mathrm{in}}\rangle $ defined in Eq. (\ref{eq:aug23_10}). The steady-state output
field state is 
\begin{equation}
\rho _{\mathrm{out}}=\left\langle \mathfrak{b},\psi \right\rangle
\left\langle \phi \right\vert \rho _{\mathrm{\infty }}|\phi \rangle
\left\langle \mathfrak{b},\psi \right\rangle ^{\ast },
\label{eq:ss_nonpassive}
\end{equation}%
where $\left\langle \mathfrak{b},\psi \right\rangle $ is given in Eq. (%
\ref{eq:inner}), and 
\begin{equation}
\rho _{\mathrm{\infty }}
\triangleq
\lim_{t_0\rightarrow-\infty, 
t\rightarrow\infty} U(t, t_{0})|\phi 0^{\otimes m}\rangle
\left\langle \phi 0^{\otimes m}\right\vert U(t, t_{0})^{\ast }
\label{eq:rho_jan15}
\end{equation}
 is a zero-mean Gaussian
state for the joint system whose power spectral density matrix is given by Eq. (\ref
{eq:temp_gaussian}) in Lemma \ref{lem:gaussian}.
\end{theorem}

\textbf{Proof. ~} When the system is initialized in the vacuum state $|\phi\rangle$ and is driven by a vacuum field $|0^{\otimes m}\rangle$, it is well-known that the steady-state joint system-field state $\rho_{\infty}$ in Eq. (\ref{eq:rho_jan15}) is a zero-mean Gaussian state whose power spectral density matrix is given by Eq. (\ref
{eq:temp_gaussian}) in Lemma \ref{lem:gaussian}, see, e.g., \cite[Chapter 6]{WM10}.   Let the input state be the $m$-channel $m$-photon input state $|\Psi _{\mathrm{in}}\rangle$ defined in Eq. (\ref{eq:aug23_10}). In this case, in steady state, the joint system-field state is 
\begin{equation}
\rho _{\rm ss} \triangleq \lim_{t_0\rightarrow-\infty, 
t\rightarrow\infty}U(t, t_{0})|\phi \Psi _{\mathrm{in}}\rangle
\left\langle \phi \Psi _{\mathrm{in}}\right\vert U(t, t_{0})^{\ast }.
\label{eq:temp_sept16_1}
\end{equation}%
The steady-state output field state, denoted by $\rho _{\mathrm{out}}$, is obtained by taking the partial trace
of $\rho _{\rm ss}$ with respect to the system, that is, 
\begin{equation}
\rho _{\mathrm{out}} = \langle\phi| \rho _{\rm ss} |\phi\rangle  =\lim_{t_0\rightarrow-\infty, 
t\rightarrow\infty}\left\langle \phi \right\vert
U(t, t_{0})|\phi \Psi _{\mathrm{in}}\rangle \left\langle \phi \Psi _{\mathrm{%
in}}\right\vert U(t, t_{0})^{\ast }|\phi \rangle .  \label{eq:temp7}
\end{equation}%
Because the system is asymptotically stable, according to Eq. (\ref{eq:out_tf_3}),%
\begin{equation}
\breve{b}_{\mathrm{in}}(t)=g_{G^{-1}}\circledast \breve{b}_{\mathrm{out}%
}(t)=g_{G^{-1}}\circledast U(t,-\infty)^{\ast }\breve{b}_{\mathrm{in}}(t)U(t,-\infty).
\label{july31_7}
\end{equation}%
Pre-multiplying Eq. (\ref{july31_7}) by $U(t,-\infty)$  and post-multiplying Eq. (\ref{july31_7}) by $U(t,-\infty)^{\ast }$, we get
\begin{equation}
U(t,-\infty)\breve{b}_{\mathrm{in}}(t)U(t,-\infty)^{\ast }=g_{G^{-1}}\circledast \breve{%
b}_{\mathrm{in}}(t).  \label{july31_6}
\end{equation}%
This, together with Eq. (\ref{eq:G_inv}), yields%
\begin{eqnarray}
&&
\lim_{t_0\rightarrow-\infty, 
t\rightarrow\infty}U(t, t_{0})|\phi \Psi _{\mathrm{in}}\rangle
\nonumber
\\
&=&
\lim_{t_0\rightarrow-\infty, 
t\rightarrow\infty}\int dt_{1}\cdots dt_{m}\ \psi _{\mathrm{in}%
}(t_{1},\ldots ,t_{m})U(t, t_{0})b_{\mathrm{in,}1}^{\ast }(t_{1})\cdots b_{%
\mathrm{in,}m}^{\ast }(t_{m})|\phi 0^{\otimes m}\rangle
\nonumber
 \\
&=&
\lim_{t_0\rightarrow-\infty, 
t\rightarrow\infty}\int dt_{1}\cdots dt_{m}\ \psi _{\mathrm{in}}(t_{1},\ldots ,t_{m})U(t, t_{0})b_{\mathrm{in,}1}^{\ast
}(t_{1})U(t, t_{0})^{\ast }\cdots U(t, t_{0})b_{\mathrm{in,}m}^{\ast
}(t_{m})U(t, t_{0})^{\ast }U(t, t_{0})|\phi 0^{\otimes m}\rangle  
\nonumber 
\\
&=&
\lim_{t\rightarrow \infty}\int_{-\infty}^{t}\cdots \int_{-\infty}^{t}dt_{1}\cdots dt_{m}\ \psi _{\mathrm{
in}}(t_{1},\ldots ,t_{m})\left\langle \phi \right\vert U(t_{1},-\infty)b_{
\mathrm{in,1}}^{\ast }(t_{1})U(t_{1},-\infty)^{\ast }  
\nonumber \\
&&
\ \ \cdots U(t_{m},-\infty)b_{\mathrm{in},m}^{\ast }(t_{m})U(t_{m},-\infty)^{\ast}\lim_{t_{0}\rightarrow -\infty ,t\rightarrow \infty}U(t,t_{0})|\phi 0^{\otimes m}\rangle  
\nonumber \\
&=&\int dt_{1}\cdots dt_{m}\;\psi _{\mathrm{in}}(t_{1},\ldots
,t_{m})\int_{-\infty }^{\infty }\left[ 
\begin{array}{cc}
-g_{G^{+}}^{1}(r_{1}-t_{1})^{\dagger } & g_{G^{-}}^{1}(r_{1}-t)^{T}
\end{array}
\right] \breve{b}_{\mathrm{in}}(r_{1})dr_{1}  
\nonumber
 \\ 
&&
\times \cdots  
\nonumber
 \\
&&
\times \int_{-\infty }^{\infty }\left[ 
\begin{array}{cc}
-g_{G^{+}}^{m}(r_{m}-t_{m})^{\dagger } & g_{G^{-}}^{m}(r_{m}-t_{m})^{T}
\end{array}
\right] \breve{b}_{\mathrm{in}}(r_{m})dr_{m} \lim_{t_{0}\rightarrow -\infty ,t\rightarrow \infty
} U(t, t_{0})|\phi 0^{\otimes
m}\rangle  
\label{jan9_1} 
\\
&=&
\int dt_{1}\cdots dt_{m}\;\psi _{\mathrm{in}}(t_{1},\ldots ,t_{m})\left\{
\sum_{j_{1}=1}^{m}\int_{-\infty }^{\infty
}-g_{G^{+}}^{j_{1}1}(r_{1}-t_{1})^{\ast }b_{\mathrm{in,}j_{1}}(r_{1})dr_{1}+
\sum_{j_{1}=1}^{m}\int_{-\infty }^{\infty }g_{G^{-}}^{j_{1}1}(r_{1}-t_{1})b_{
\mathrm{in,}j_{1}}^{\ast }(r_{1})dr_{1}\right\}  \nonumber \\
&&\times \cdots  \nonumber \\
&&\times \left\{ \sum_{j_{m}=1}^{m}\int_{-\infty }^{\infty
}-g_{G^{+}}^{j_{m}m}(r_{m}-t_{m})^{\ast }b_{\mathrm{in,}j_{m}}(r_{m})dr_{m}+
\sum_{j_{m}=1}^{m}\int_{-\infty }^{\infty }g_{G^{-}}^{j_{m}m}(r_{m}-t_{m})b_{
\mathrm{in,}j_{m}}^{\ast }(r_{m})dr_{m}\right\}
\nonumber
\\
&& \times 
 \lim_{t_{0}\rightarrow -\infty ,t\rightarrow \infty
} U(t, t_{0})|\phi 0^{\otimes
m}\rangle  
\nonumber
 \\
&=&
\sum_{j_{1},\cdots ,j_{m}=1}^{m}\sum_{d_{1},\cdots ,d_{m}=\pm 1}\int
dr_{1}\cdots dr_{m}\quad \psi _{j_{1}\cdots j_{m}}^{d_{1}\cdots
d_{m}}(r_{1},\ldots ,r_{m})b_{j_{1}}^{d_{1}}(r_{1})\cdots
b_{j_{m}}^{d_{m}}(r_{m})\lim_{t_{0}\rightarrow -\infty ,t\rightarrow \infty
} U(t, t_{0})|\phi 0^{\otimes m}\rangle  
\nonumber
 \\
&=&
\left\langle \mathfrak{b},\psi \right\rangle \lim_{t_{0}\rightarrow -\infty ,t\rightarrow \infty
} U(t, t_{0})|\phi 0^{\otimes
m}\rangle .  
\label{july31_8}
\end{eqnarray}
Notice that in Eq. (\ref{jan9_1}), $g_{G^\pm}^j$ stands for the $j$th column of $g_{G^\pm}$ for $j=1,\ldots,m$. Substituting Eq. (\ref{july31_8}) into Eq. (\ref{eq:temp7}) we obtain 
\begin{eqnarray}
\rho _{\mathrm{out}} 
&=&
\lim_{t_0\rightarrow-\infty, 
t\rightarrow\infty}\left\langle \phi \right\vert
U(t, t_{0})|\phi \Psi _{\mathrm{in}}\rangle \left\langle \phi \Psi _{\mathrm{in}}\right\vert U(t, t_{0})^{\ast }|\phi \rangle  
\nonumber
 \\
&=&
\left\langle \mathfrak{b},\psi \right\rangle \ \lim_{t_0\rightarrow-\infty, 
t\rightarrow\infty}\left\langle \phi \right\vert U(t, t_{0})|\phi 0^{\otimes m}\rangle
\left\langle \phi 0^{\otimes m}\right\vert U(t, t_{0})^{\ast }|\phi \rangle
\left\langle \mathfrak{b},\psi \right\rangle ^{\ast }  
\nonumber 
\\
&=&
\left\langle \mathfrak{b},\psi \right\rangle \left\langle \phi
\right\vert \rho _{\mathrm{\infty }}|\phi \rangle \left\langle \mathfrak{b},\psi \right\rangle ^{\ast },  
\label{july31_9}
\end{eqnarray}
which is exactly Eq. (\ref{eq:ss_nonpassive}). $\blacksquare $

\section{$N$ ($N\geq m$) photons superposed over $m$ channels}\label{sec:mn}

In this section, we study how a \textit{passive} quantum linear system
processes $N$ photons that are superposed over $m$ input channels, thus
generalizing the results in Section \ref{sec:mm}.

\subsection{State transfer}\label{sec:mn_aug14}

Let the input field be in a state where $N$ photons are superposed over $m$
input channels. Specifically, the input state considered in this subsection
is defined to be 
\begin{equation}
|\Psi _{\mathrm{in}}\rangle 
\triangleq
\int d\overrightarrow{t}\;\psi _{\mathrm{in}}(t_{1}^{1},\ldots
,t_{k_{1}}^{1},\ldots ,t_{1}^{m},\ldots ,t_{k_{m}}^{m}) \ b_{\mathrm{in}
,1}^{\ast }(t_{1}^{1})\cdots b_{\mathrm{in},1}^{\ast }(t_{k_{1}}^{1})\cdots
b_{\mathrm{in},m}^{\ast }\left( t_{1}^{m}\right) \cdots b_{\mathrm{in}
,m}^{\ast }\left( t_{k_{m}}^{m}\right) |0^{\otimes m}\rangle .
\label{eq:aug16_5}
\end{equation}
In Eq. (\ref{eq:aug16_5}), the positive integers $k_{i}$ satisfy $
\sum_{i=1}^{m}k_{i}=N$. It is also implicitly assumed in Eq. (\ref{eq:aug16_5}) that the $N$-variable function $\psi_{\rm in}$ normalizes the state $|\Psi _{\mathrm{in}}\rangle$.

\begin{remark}\label{rem:dec27_1}
 For the $i$th input channel $(i=1,\ldots ,m)$, the creation
operator $b_{\mathrm{in},i}^{\ast }$ appears $k_{i}$ times in Eq. (\ref{eq:aug16_5}), thus there are $k_{i}$ photons in the $i$th input channel.
\end{remark}

With the notation introduced in Eq. (\ref{dec3_dot}), Eq. (\ref{eq:aug16_5}) can be re-written as 
\begin{equation}
|\Psi _{\mathrm{in}}\rangle 
=
\psi_{\rm in} \cdot
_{k_{1}\cdots k_{m}}^{N}b_{\mathrm{in}}^{\#}|0^{\otimes m}\rangle.
\label{dec11_4}
\end{equation}
Moreover, inspired by Eq. (\ref{dec_up2}), update the $N$-variable function $\psi _{\mathrm{in}}(t_{1}^{1},\ldots
,t_{k_{1}}^{1},\ldots ,t_{1}^{m},\ldots ,t_{k_{m}}^{m})$ to an $N$-way $m$-dimensional tensor $\psi _{\mathrm{in}}^{\uparrow }(t_{1}^{1},\ldots
,t_{k_{1}}^{1},\ldots ,t_{1}^{m},\ldots ,t_{k_{m}}^{m})$, whose elements are
defined as 
\begin{eqnarray}
&&\psi _{\mathrm{in,}i_{1}^{1}\ldots i_{k_{1}}^{1}\ldots i_{1}^{m}\ldots
i_{k_{m}}^{m}}^{\uparrow }(t_{1}^{1},\ldots ,t_{k_{1}}^{1},\ldots
,t_{1}^{m},\ldots ,t_{k_{m}}^{m})  \nonumber \\
&\triangleq &\left\{ 
\begin{array}{ll}
\psi _{\mathrm{in}}(t_{1}^{1},\ldots ,t_{k_{1}}^{1},\ldots ,t_{1}^{m},\ldots
,t_{k_{m}}^{m}), & \mathrm{if\ }i_{1}^{1}=1,\cdots
,i_{k_{1}}^{1}=k_{1},\cdots ,i_{1}^{m}=\sum\limits_{j=1}^{m-1}k_{j}+1,\cdots
,i_{k_{m}}^{m}=N, \\ 
0, & \mathrm{otherwise}.
\end{array}
\right.   \label{aug1_14b}
\end{eqnarray}
Then, Eq. (\ref{dec11_4}) can be re-written as
\begin{equation}
|\Psi _{\mathrm{in}}\rangle 
=
\psi _{\mathrm{in}}^{\uparrow } \odot _{k_{1}\cdots k_{m}}^{N}b_{\mathrm{in}}^{\#}|0^{\otimes m}\rangle ,
\label{jan9_psi_out}
\end{equation}
where the operation $\odot _{k_{1}\cdots k_{m}}^{N}$ has been introduced in Eq. (\ref{dec3_dot^2}).

The following result gives an explicit form of the steady-state output field state.

\begin{theorem}\label{thm:aug23}

If the asymptotically stable \textit{passive} quantum linear system (\ref
{sys_passive}) is initialized in the vacuum state and is driven by the $m$-channel $N$-photon input state $|\Psi
_{\mathrm{in}}\rangle $ defined in Eq. (\ref{eq:aug16_5}), then the steady-state output field state is
\begin{equation}
|\Psi _{\mathrm{out}}\rangle 
=
\psi _{\mathrm{out}}\odot _{k_{1}\cdots k_{m}}^{N}b_{\mathrm{in}}^{\#}|0^{\otimes m}\rangle 
\label{eq:aug23_2}
\end{equation}
with the pulse shape  
\begin{equation}
\psi _{\mathrm{out}}
=
\psi _{\mathrm{in}}^{\uparrow } \circledast
_{t, k_{1}\cdots k_{m}}^{N}g_{G^{-}}.  
\label{eq:aug23_1}
\end{equation}
In Eq. (\ref{eq:aug23_1}), the tensor $\psi _{\mathrm{in}}^{\uparrow }$ has been given in Eq. (\ref{aug1_14b}),  and the operation $\circledast
_{t, k_{1}\cdots k_{m}}^{N}$ has been defined in Eq. (\ref{dec11_2}).
\end{theorem}

\textbf{Proof. ~} The proof is similar to that of Theorem \ref{thm:passive_mm}, so is omitted.

\begin{example} \label{ex:BS}
 Consider a beamsplitter
\begin{equation}
S_{-}=\left[ 
\begin{array}{cc}
R & -T \\ 
T & R
\end{array}
\right] ,~~R,T\in \mathbb{C},~|R|^{2}+|T|^{2}=1,  \label{aug1_temp3}
\end{equation}
and a 3-photon input state of the form 
\begin{equation}
|\Psi _{\mathrm{in}}\rangle 
=
\int_{-\infty
}^{\infty }\int_{-\infty }^{\infty }\int_{-\infty }^{\infty
}dt_{1}dt_{2}dt_{3}\;\psi _{\mathrm{in}}(t_{1},t_{2},t_{3})\;b_{\mathrm{in,}1}^{\ast }(t_{1})b_{\mathrm{in,}2}^{\ast }(t_{2})b_{\mathrm{in,}2}^{\ast
}(t_{3})|0_{1}\rangle \otimes |0_{2}\rangle .  \label{eq:aug21_14}
\end{equation}
In this case, there are two input channels ($m=2$). The total number of
photons is $N=3$. In fact, as explained in Remark \ref{rem:dec27_1} above, there is one photon in the first channel ($k_{1}=1$) and two photons in the second channel ($k_{2}=2$). Simple
calculation yields
\begin{equation}
\left\langle \Psi _{\mathrm{in}}|\Psi _{\mathrm{in}}\right\rangle 
=
\int_{-\infty }^{\infty }\int_{-\infty }^{\infty
}\int_{-\infty }^{\infty }dt_{1}dt_{2}dt_{3}\;\left( \psi _{\mathrm{in}
}^{\ast }(t_{1},t_{3},t_{2})\psi _{\mathrm{in}}(t_{1},t_{2},t_{3})+\left
\vert \psi _{\mathrm{in}}(t_{1},t_{2},t_{3})\right\vert ^{2}\right) .
\label{aug1_14}
\end{equation}
Therefore, the normalization condition requires that
\begin{equation}
\int_{-\infty }^{\infty }\int_{-\infty }^{\infty
}\int_{-\infty }^{\infty }dt_{1}dt_{2}dt_{3}\;\left( \psi _{\mathrm{in}
}^{\ast }(t_{1},t_{3},t_{2})\psi _{\mathrm{in}}(t_{1},t_{2},t_{3})+\left
\vert \psi _{\mathrm{in}}(t_{1},t_{2},t_{3})\right\vert ^{2}\right)
=
1 .
\label{july23_2}
\end{equation}
According to Theorem \ref{thm:aug23}, the steady-state output field state is
\begin{eqnarray}
&&
|\Psi _{\mathrm{out}}\rangle  
\nonumber \\
&=&
RT^{2}\int d\overrightarrow{t}\;b_{\mathrm{in},1}^{\ast }(t_{1})b_{\mathrm{in},1}^{\ast }(t_{2})b_{\mathrm{in},1}^{\ast }(t_{3})\psi _{\mathrm{in}}(t_{1},t_{2},t_{3})|0_{1}\rangle
\otimes |0_{2}\rangle  
\nonumber \\
&&
-T\int d\overrightarrow{t}\;b_{\mathrm{in}
,1}^{\ast }(t_{1})b_{\mathrm{in},1}^{\ast }(t_{2})b_{\mathrm{in},2}^{\ast
}(t_{3})\left[ R^{2}\psi _{\mathrm{in}}(t_{1},t_{2},t_{3})+R^{2}\psi _{
\mathrm{in}}(t_{1},t_{3},t_{2})-T^{2}\psi _{\mathrm{in}}(t_{3},t_{2},t_{1})
\right] |0_{1}\rangle \otimes |0_{2}\rangle  
\nonumber \\
&&
-R\int d\overrightarrow{t}\;b_{\mathrm{in},1}^{\ast }(t_{1})b_{\mathrm{in},2}^{\ast }(t_{2})b_{\mathrm{in},2}^{\ast
}(t_{3})\left[ T^{2}\psi _{\mathrm{in}}(t_{2},t_{1},t_{3})+T^{2}\psi _{
\mathrm{in}}(t_{3},t_{2},t_{1})-R^{2}\psi _{\mathrm{in}}(t_{1},t_{2},t_{3})
\right] |0_{1}\rangle \otimes |0_{2}\rangle  
\nonumber \\
&&
+|0_{1}\rangle
\otimes R^{2}T\int d\overrightarrow{t}\;b_{\mathrm{in},2}^{\ast }(t_{1})b_{\mathrm{in},2}^{\ast }(t_{2})b_{\mathrm{in},2}^{\ast }(t_{3})\psi _{\mathrm{in}}(t_{1},t_{2},t_{3}) |0_{2}\rangle .  
\label{july23_1}
\end{eqnarray}
In particular, if $R=T=\frac{1}{\sqrt{2}}$, Eq. (\ref{july23_1}) reduces to 
\begin{eqnarray}
&&
|\Psi _{\mathrm{out}}\rangle  
\nonumber \\
&=&
\frac{1}{2\sqrt{2}}\int d\overrightarrow{t}\;b_{\mathrm{in},1}^{\ast }(t_{1})b_{\mathrm{in},1}^{\ast }(t_{2})b_{\mathrm{in},1}^{\ast
}(t_{3})\psi _{\mathrm{in}}(t_{1},t_{2},t_{3})|0_{1}\rangle \otimes
|0_{2}\rangle  
\nonumber \\
&&
-\frac{1}{2\sqrt{2}}\int d\overrightarrow{t}\;b_{\mathrm{in},1}^{\ast }(t_{1})b_{\mathrm{in},1}^{\ast }(t_{2})b_{\mathrm{in},2}^{\ast
}(t_{3})\left[ \psi _{\mathrm{in}}(t_{1},t_{2},t_{3})+\psi _{\mathrm{in}}(t_{1},t_{3},t_{2})-\psi _{\mathrm{in}}(t_{3},t_{2},t_{1})\right]
|0_{1}\rangle \otimes |0_{2}\rangle  
\nonumber \\
&&
-\frac{1}{2\sqrt{2}}\int d\overrightarrow{t}\;b_{\mathrm{in},1}^{\ast }(t_{1})b_{\mathrm{in},2}^{\ast }(t_{2})b_{\mathrm{in},2}^{\ast
}(t_{3})\left[ \psi _{\mathrm{in}}(t_{2},t_{1},t_{3})+\psi _{\mathrm{in}}(t_{3},t_{2},t_{1})-\psi _{\mathrm{in}}(t_{1},t_{2},t_{3})\right]
|0_{1}\rangle \otimes |0_{2}\rangle  
\nonumber \\
&&
+|0_{1}\rangle \otimes \frac{1}{2\sqrt{2}}\int d\overrightarrow{t}\;b_{\mathrm{in},2}^{\ast }(t_{1})b_{\mathrm{in},2}^{\ast }(t_{2})b_{\mathrm{in},2}^{\ast
}(t_{3})\psi _{\mathrm{in}}(t_{1},t_{2},t_{3})
|0_{2}\rangle .  
\label{july18_2}
\end{eqnarray}
Furthermore, if $\psi _{\mathrm{in}}(t_{1},t_{2},t_{3})$ is 
permutation-invariant, and 
\begin{equation}
\left\Vert \psi _{\mathrm{in}}(t_{1},t_{2},t_{3})
\right\Vert 
=
 \sqrt{\int_{-\infty }^{\infty }\int_{-\infty }^{\infty
}\int_{-\infty }^{\infty }dt_{1}dt_{2}dt_{3}\  |\psi_{\rm in} (t_1,t_2,t_3)|^2}=\frac{1}{\sqrt{2}},
\end{equation}
 then by Eq. (\ref{july23_2}), the state $|\Psi_{\rm in}\rangle$ is normalized. Furthermore,  Eq. (\ref{july18_2}) becomes
\begin{eqnarray}
&&|\Psi _{\mathrm{out}}\rangle  \nonumber \\
&=&\frac{1}{2\sqrt{2}}\int d\overrightarrow{t}\;b_{\mathrm{in},1}^{\ast }(t_{1})b_{
\mathrm{in},1}^{\ast }(t_{2})b_{\mathrm{in},1}^{\ast }(t_{3})\psi _{
\mathrm{in}}(t_{1},t_{2},t_{3})|0_{1}\rangle \otimes |0_{2}\rangle  \nonumber \\
&&-\frac{1}{2\sqrt{2}}\int d\overrightarrow{t}\;b_{\mathrm{in},1}^{\ast }(t_{1})b_{
\mathrm{in},1}^{\ast }(t_{2})b_{\mathrm{in},2}^{\ast }(t_{3})\psi _{
\mathrm{in}}(t_{1},t_{2},t_{3})|0_{1}\rangle \otimes |0_{2}\rangle  \nonumber \\
&&-\frac{1}{2\sqrt{2}}\int d\overrightarrow{t}\;b_{\mathrm{in},1}^{\ast }(t_{1})b_{
\mathrm{in},2}^{\ast }(t_{2})b_{\mathrm{in},2}^{\ast }(t_{3})\psi _{
\mathrm{in}}(t_{1},t_{2},t_{3})|0_{1}\rangle \otimes |0_{2}\rangle  \nonumber \\
&&+|0_{1}\rangle \otimes\frac{1}{2\sqrt{2}}\int d\overrightarrow{t}\;b_{\mathrm{in},2}^{\ast }(t_{1})b_{
\mathrm{in},2}^{\ast }(t_{2})b_{\mathrm{in},2}^{\ast }(t_{3})\psi _{
\mathrm{in}}(t_{1},t_{2},t_{3}) |0_{2}\rangle .
\label{july18_3}
\end{eqnarray}
Define states 
\begin{eqnarray}
\left\vert \Pi _{30}\right\rangle
&\triangleq &
\frac{1}{\sqrt{3}}\int d\overrightarrow{t}\;b_{\mathrm{in},1}^{\ast }(t_{1})b_{\mathrm{in},1}^{\ast }(t_{2})b_{\mathrm{in},1}^{\ast }(t_{3})\psi _{\mathrm{in}}(t_{1},t_{2},t_{3})|0_{1}\rangle ,  
\nonumber \\
\left\vert \Pi _{21}\right\rangle 
&\triangleq &
\int d
\overrightarrow{t}\;b_{\mathrm{in},1}^{\ast }(t_{1})b_{\mathrm{in},1}^{\ast }(t_{2})b_{\mathrm{in},2}^{\ast }(t_{3})\psi _{\mathrm{in}}(t_{1},t_{2},t_{3})|0_{1}\rangle \otimes |0_{2}\rangle ,  
\nonumber \\
\left\vert \Pi _{12}\right\rangle 
&\triangleq &
\int d
\overrightarrow{t}\;b_{\mathrm{in},1}^{\ast }(t_{1})b_{\mathrm{in},2}^{\ast }(t_{2})b_{\mathrm{in},2}^{\ast }(t_{3})\psi _{\mathrm{in}}(t_{1},t_{2},t_{3})|0_{1}\rangle \otimes |0_{2}\rangle ,  
\nonumber \\
\left\vert \Pi _{03}\right\rangle 
&\triangleq &
\frac{1}{\sqrt{3}}\int d\overrightarrow{t}\;b_{\mathrm{in},2}^{\ast }(t_{1})b_{\mathrm{in},2}^{\ast }(t_{2})b_{\mathrm{in},2}^{\ast }(t_{3})\psi _{\mathrm{in}}(t_{1},t_{2},t_{3})|0_{2}\rangle .  
\label{july18_4}
\end{eqnarray}
It is easy to show that all the states in Eq. (\ref{july18_4}) are
normalized. Moreover, $\left\vert \Pi _{30}\right\rangle $ is a $3$-photon
state for the first channel, $\left\vert \Pi _{03}\right\rangle $ is a $3$-photon state for the second channel, and $\left\vert \Pi _{21}\right\rangle 
$ and $\left\vert \Pi _{12}\right\rangle $ are states where two channels
share three photons. With these new notations, Eq. (\ref{july18_3}) can be
re-written as
\begin{equation}
|\Psi _{\mathrm{out}}\rangle =\frac{\sqrt{6}}{4}\left\vert \Pi
_{30}\right\rangle \otimes |0_{2}\rangle -\frac{\sqrt{2}}{4}\left\vert \Pi
_{21}\right\rangle -\frac{\sqrt{2}}{4}\left\vert \Pi _{12}\right\rangle
+\left\vert 0_{1}\right\rangle \otimes \frac{\sqrt{6}}{4}\left\vert \Pi
_{03}\right\rangle .  \label{july18_5}
\end{equation}
Finally, if 
\begin{equation}
\psi _{\mathrm{in}}(t_{1},t_{2},t_{3})=\xi _{1}(t_{1})\xi _{2}(t_{2})\xi
_{3}(t_{3}),  \label{july19_4}
\end{equation}
then Eq. (\ref{eq:aug21_14}) becomes
\begin{eqnarray*}
|\Psi _{\mathrm{in}}\rangle 
&=&
\int_{-\infty
}^{\infty }\xi _{1}(t_{1})b_{\mathrm{in,}1}^{\ast
}(t_{1})dt_{1}|0_{1}\rangle \otimes \int_{-\infty }^{\infty }\int_{-\infty
}^{\infty }dt_{2}dt_{3}\;\xi _{2}(t_{2})\xi _{3}(t_{3})\;b_{\mathrm{in,}2}^{\ast }(t_{2})b_{\mathrm{in,}2}^{\ast }(t_{3})|0_{2}\rangle 
\\
&=&
\mathbf{B}_{\mathrm{in,}1}^{\ast }(\xi
_{1})|0_{1}\rangle \otimes \mathbf{B}_{\mathrm{in,}2}^{\ast }(\xi _{2})
\mathbf{B}_{\mathrm{in,}2}^{\ast }(\xi _{3})|0_{3}\rangle .
\end{eqnarray*}
That is, the input is a product state, with one photon in channel 1 and two
photons in channel 2. Moreover, if $\xi_1= \xi_2 = \xi_3 \equiv \xi$, then the normalization condition requires that $\|\xi\|=\frac{1}{\sqrt[6]{2}}$.  In this case, Eq. (\ref{july18_4}) reduces to 
\begin{eqnarray}
\left\vert \Pi _{30}\right\rangle 
&=&
\frac{1}{\sqrt{3}}(\mathbf{B}_{\mathrm{in,}1}^{\ast }(\xi))^3|0_{1}\rangle ,  
\nonumber \\
\left\vert \Pi _{21}\right\rangle
 &=&
(\mathbf{B}_{\mathrm{in,}1}^{\ast }(\xi))^2 |0_{1}\rangle \otimes \mathbf{B}_{\mathrm{in,}2}^{\ast }(\xi)|0_{2}\rangle , 
 \nonumber \\
\left\vert \Pi _{12}\right\rangle 
&=&
\mathbf{B}_{\mathrm{in,}1}^{\ast }(\xi)|0_{1}\rangle \otimes (\mathbf{B}_{\mathrm{in,}
2}^{\ast }(\xi))^2 |0_{2}\rangle , 
 \nonumber \\
\left\vert \Pi _{03}\right\rangle 
&=&
\frac{1}{\sqrt{3}} 
(\mathbf{B}_{\mathrm{in,}2}^{\ast }(\xi))^3 |0_{2}\rangle . 
 \label{july31_10}
\end{eqnarray}
That is, all the states become product states. If we ignore pulse shapes and only count the number of photons in each channel, we
may identify $\left\vert \Pi _{30}\right\rangle $ with $\left\vert
3_{1}\right\rangle $, $\left\vert \Pi _{03}\right\rangle $ with $\left\vert
3_{2}\right\rangle $, $\left\vert \Pi _{21}\right\rangle $ with $\left\vert
2_{1}\right\rangle \otimes \left\vert 1_{2}\right\rangle $, and $\left\vert
\Pi _{12}\right\rangle $ with $\left\vert 1_{1}\right\rangle \otimes
\left\vert 2_{2}\right\rangle $. Accordingly, the state in Eq. (\ref{july18_5}) reduces to 
\begin{equation}
|\Psi _{\mathrm{out}}\rangle =\frac{\sqrt{6}}{4}\left\vert
3_{1}\right\rangle \otimes \left\vert 0_{2}\right\rangle -\frac{\sqrt{2}}{4}
\left\vert 2_{1}\right\rangle \otimes \left\vert 1_{2}\right\rangle -\frac{
\sqrt{2}}{4}\left\vert 1_{1}\right\rangle \otimes \left\vert
2_{2}\right\rangle +\frac{\sqrt{6}}{4}\left\vert 0_{1}\right\rangle \otimes
\left\vert 3_{2}\right\rangle .  
\label{july19_6}
\end{equation}
\end{example}


\subsection{The invariant set}\label{sec:invariant_aug3}

In this subsection, we define a class of $m$-channel $N$-photon states and show that this
class of states is invariant under the steady-state action of a quantum
linear passive system. The discussions here generalize those in
Subsection \ref{subsec:invariant}.

Motivated by Eqs. (\ref{jan9_psi_out}) and (\ref{eq:aug23_2}), define a class of $m$-channel $N$-photon states:
\begin{equation}
\mathcal{F}_{2}\triangleq \left\{ |\Psi \rangle =\psi \odot _{k_{1}\cdots
k_{m}}^{N}b_{\mathrm{in}}^{\#}|0^{\otimes m}\rangle \ |\mathrm{the~}N\text{\textrm{-}}\mathrm{way}~m\text{\textrm{-}}\mathrm{dimensional~tensor~function~}\psi \mathrm{~normalizes~}|\Psi \rangle .\right\}  
\label{july19_8}
\end{equation}

The following result shows that the set $\mathcal{F}_{2}$ is invariant under
the steady-state action of a passive quantum linear system.


\begin{theorem}
\label{thm:passive_invariance_general} 
The steady-state output field state of the asymptotically stable 
passive quantum linear system (\ref{sys_passive}), initialized in the vacuum state $|\phi
\rangle $ and driven by an $m$-channel $N$-photon input state $|\Psi _{\mathrm{in}}\rangle \in 
\mathcal{F}_{2}$ with pulse information encoded by an $N$-way $m$-dimensional tensor function $\psi _{\mathrm{in}}$, is another element $|\Psi
_{\mathrm{out}}\rangle \in \mathcal{F}_{2}$, whose pulse information is
encoded by an $N$-way $m$-dimensional tensor function $\psi _{\mathrm{out}}$
given by 
\begin{equation}
\psi _{\mathrm{out}}=\psi _{\mathrm{in}}\circledast _{t, k_{1}\cdots
k_{m}}^{N}g_{G^{-}}.  \label{eq:mar22_1}
\end{equation}
\end{theorem}

This result can be established in a similar way as Theorem \ref{thm:passive_invariance}. So the proof is omitted.


\section{An arbitrary number of photons superposed over $m$ input channels}

\label{subsec:mN_general}

In all the previous discussions, we have implicitly assumed that the total
number of photons is no less than the number of input channels. In this
section, we remove this constraint. More specifically, we study a class of $m
$-channel $N$-photon states where $N$ can be an arbitrary positive integer.


\subsection{A class of $m$-channel $N$-photon input states}\label{subsec:mN_input}

In this subsection, we present a class of $m$-channel $N$-photon input
states. Two illustrative examples are also given.

Let a normalized $m$-channel $N$-photon input state be 
\begin{equation}
|\Psi _{\mathrm{in}}\rangle 
=
\tprod\limits_{j=1}^{N}\sum_{k=1}^{m}\int_{-\infty }^{\infty }dt\ \psi _{\mathrm{in},jk}(t)b_{\mathrm{in},k}^{\ast }(t)|0^{\otimes m}\rangle ,
\label{eq:temp1}
\end{equation}
where $N$ is an arbitrary positive integer. The input state $|\Psi _{\mathrm{in}}\rangle $ is
parametrized by the pulse shapes $\psi _{\mathrm{in},jk}(t)$ ($j=1,\ldots ,N$
and $k=1,\ldots m$). Clearly, different combinations of $\psi _{\mathrm{in},jk}(t)$ give rise to different $m$-channel $N$-photon states. By the
notation in Eq. (\ref{july31_B}), the $m$-channel $N$-photon input state in
Eq. (\ref{eq:temp1}) can be re-written as 
\begin{equation}
|\Psi _{\mathrm{in}}\rangle 
=
\tprod\limits_{j=1}^{N}\sum_{k=1}^{m}\mathbf{B}_{\mathrm{in},k}^{\ast }(\psi
_{\mathrm{in},jk})|0^{\otimes m}\rangle .  \label{eq:temp1b}
\end{equation}

\begin{remark}
\label{rem:zhang13} A class of photon-Gaussian states has been defined in 
\cite[Eq. (95)]{zhang13}. If the density matrix $\rho _{R}$ there used is of
the form $\rho _{R}=|\phi 0^{\otimes m}\rangle \left\langle \phi 0^{\otimes
m}\right\vert $, and moreover, $\xi _{jk}^{+}\equiv 0$, then the resulting
states are $m$-channel $m$-photon states. Actually, they form a special
subclass of the $m$-channel $N$-photon states defined in Eq. (\ref{eq:temp1b})
(with $N=m$).
\end{remark}

\begin{remark} \label{rem:jan15_photon}
Although the positive integer $N$ in Eq. (\ref{eq:temp1}) is allowed to be arbitrary, the multi-photon input states defined in Eq. (\ref{eq:temp1})  may not be able to include those multi-photon states studied in Sections \ref{sec:mm} and \ref{sec:mn} as subclasses. This can be easily seen by comparing the forms of multi-photon states in Eqs. (\ref{eq:aug23_10}), (\ref{eq:aug16_5}),  and  (\ref{eq:temp1}). 
\end{remark}

\begin{remark} \label{rem:N}
 Eq. (\ref{eq:temp1b}) provides flexibility for specifying multi-channel multi-photon states. 
 \begin{description}
 \item[(i)] if for some $j_{0}$ ($1\leq j_{0}\leq N$) and $k_{0}$ ($1\leq k_{0}\leq m$), $\psi _{\mathrm{in},j_{0}k_{0}}\equiv 0$, then the term $\mathbf{B}_{\mathrm{in},k_{0}}^{\ast }(\psi _{\mathrm{in},j_{0}k_{0}})$ does not appear on the right-hand side of Eq. (\ref{eq:temp1b}).
 
\item[(ii)]  As a special case of item (i) above, if for some $j_{0}$ ($1\leq j_{0}\leq N$), $\psi _{\mathrm{in},j_{0}k}\equiv 0$ for all $k=1,\ldots m$, then Eq. (\ref{eq:temp1b}) reduces to 
\begin{equation*}
|\Psi _{\mathrm{in}}\rangle 
=
\tprod\limits_{j=1,j\neq j_{0}}^{N}\sum_{k=1}^{m}\mathbf{B}_{\mathrm{in},k}^{\ast }(\psi _{\mathrm{in},jk})|0^{\otimes m}\rangle .
\end{equation*}
In this case, there are $\mathbf{N-1}$ photons among $m$ channels.  Thus, the term  ``$N$-photon'' is a bit confusing. Nevertheless, the exact number of photons can be determined  easily from the context.
\end{description}
\end{remark}

We illustrate Remark \ref{rem:N} with the following two Examples.

\begin{example}\label{ex:12} 
When $N=1$ and $m=2$, by Eq. (\ref{eq:temp1b}), the input
state is
\begin{equation}
|\Psi _{\mathrm{in}}\rangle 
=
\mathbf{B}_{\mathrm{in},1}^{\ast }(\psi _{\mathrm{in},11})|0_{1}\rangle \otimes
|0_{2}\rangle +|0_{1}\rangle \otimes \mathbf{B}_{\mathrm{in},2}^{\ast }(\psi _{\mathrm{in},12})|0_{2}\rangle .
\label{july19_1b}
\end{equation}
 In what follows, we discuss two cases.

Case 1): $\psi _{\mathrm{in},11}\equiv 0$. In this case, as commented by  item (i) in Remark \ref{rem:N}, Eq. (\ref{july19_1b}) becomes 
\begin{equation}
|\Psi _{\mathrm{in}}\rangle =|0_{1}\rangle \otimes \mathbf{B}_{\mathrm{in},2}^{\ast }(\psi _{\mathrm{in},12})|0_{2}\rangle .  
\label{july19_1c}
\end{equation}
The normalization condition is 
\begin{equation*}
1=\left\langle \Psi _{\mathrm{in}}|\Psi _{\mathrm{in}}\right\rangle =\int_{-\infty }^{\infty }\left\vert \psi _{\mathrm{in},12}(t)\right\vert ^{2}dt.
\end{equation*}
 In this case, the first channel is in the
vacuum state and the second channel is in a single-photon state.

Case 2): $\psi _{\mathrm{in},11}=\psi _{\mathrm{in},12}\equiv \xi $. Eq. (\ref{july19_1b}) becomes 
\begin{equation}
|\Psi _{\mathrm{in}}\rangle =\boldsymbol{B}_{\mathrm{in,}1}^{\ast }(\xi )\left\vert 0_{1}\right\rangle \otimes \left\vert
0_{2}\right\rangle +\left\vert 0_{1}\right\rangle \otimes 
\boldsymbol{B}_{\mathrm{in,}2}^{\ast }(\xi )\left\vert 0_{2}\right\rangle .
\label{aug23_1}
\end{equation}
The normalization condition
\begin{equation*}
1=\left\langle \Psi _{\mathrm{in}}|\Psi _{\mathrm{in}}\right\rangle =2\int_{-\infty }^{\infty }\left\vert \xi (t)\right\vert
^{2}dt=2\Vert \xi \Vert ^{2}
\end{equation*}
requires that $\|\xi\|=\frac{1}{\sqrt{2}}$. Moreover, it can be readily shown that 
\begin{equation}
\lim_{t_{0}\rightarrow -\infty ,t\rightarrow \infty }\left\langle \Psi _{\mathrm{in}}|\Lambda _{\mathrm{in}}(t)|\Psi _{\mathrm{in}}\right\rangle =
\frac{1}{2}\left[ 
\begin{array}{cc}
1 & 0 \\ 
0 & 1
\end{array}
\right] .  \label{aug23_1b}
\end{equation}
That is, the photon is not localized in either of the two channels; instead,
it is shared by two channels. This reveals the wave property of photons. 
\end{example}


\begin{example}
\label{ex:23} Let $N=2$ and $m=3$. According to Eq. (\ref{eq:temp1b}), the
input state is
\begin{equation}
|\Psi _{\mathrm{in}}\rangle =\sum_{k=1}^{3}
\mathbf{B}_{\mathrm{in},k}^{\ast }(\psi _{\mathrm{in},1k})\sum_{k=1}^{3}
\mathbf{B}_{\mathrm{in},k}^{\ast }(\psi _{\mathrm{in},2k})|0^{\otimes
3}\rangle .  
\label{july19_2}
\end{equation}
If $\psi _{\mathrm{in},11} = \psi _{\mathrm{in},22}\equiv 0$, then, as commented by item (i) in Remark \ref{rem:N}, the
input state in Eq. (\ref{july19_2}) becomes
\begin{eqnarray}
&&
|\Psi _{\mathrm{in}}\rangle   \nonumber \\
&=&
\mathbf{B}_{\mathrm{in},1}^{\ast }(\psi _{\mathrm{in},21})|0_{1}\rangle \otimes \mathbf{B}_{\mathrm{in},2}^{\ast
}(\psi _{\mathrm{in},12})|0_{2}\rangle \otimes |0_{3}\rangle +\mathbf{B}_{\mathrm{in},1}^{\ast }(\psi _{\mathrm{in},21})|0_{1}\rangle \otimes |0_{2}\rangle \otimes \mathbf{B}_{\mathrm{in},3}^{\ast }(\psi _{\mathrm{in},13})|0_{3}\rangle  
 \nonumber \\
&&
+|0_{1}\rangle \otimes \mathbf{B}_{\mathrm{in},2}^{\ast }(\psi _{\mathrm{in},12})|0_{2}\rangle \otimes \mathbf{B}_{\mathrm{in},3}^{\ast }(\psi _{\mathrm{in},23})|0_{3}\rangle +|0_{1}\rangle \otimes |0_{2}\rangle \otimes \mathbf{B}_{\mathrm{in},3}^{\ast }(\psi _{\mathrm{in},13})\mathbf{B}_{\mathrm{in},3}^{\ast }(\psi _{\mathrm{in},23})|0_{3}\rangle .  
\label{july19_2a}
\end{eqnarray}
That is, two photons are shared by three channels. If further $\psi _{\mathrm{in},21}\equiv \psi _{\mathrm{in},13}\equiv 0$, then, as commented by item (i) in Remark \ref{rem:N}, Eq. (\ref{july19_2a}) reduces to
\begin{equation}
|\Psi _{\mathrm{in}}\rangle =|0_{1}\rangle
\otimes \mathbf{B}_{\mathrm{in},2}^{\ast }(\psi _{\mathrm{in},12})|0_{2}\rangle \otimes \mathbf{B}_{\mathrm{in},3}^{\ast }(\psi _{\mathrm{in},23})|0_{3}\rangle .  \label{july19_2b}
\end{equation}
In this case, the first channel is in the vacuum state, and there is exactly
one photon in each of the second and third channels, respectively. Finally, if further $\psi _{\mathrm{in},23}(t)\equiv 0$, the only existing pulse shape in Eq. (\ref{july19_2}) is $\psi _{\mathrm{in},12}$, and therefore, as commented by item (ii) in
Remark \ref{rem:N}, we end up with a single photon state. Indeed, Eq. (\ref{eq:temp1b}) reduces to
\begin{equation}
|\Psi _{\mathrm{in}}\rangle =|0_{1}\rangle
\otimes \mathbf{B}_{\mathrm{in},2}^{\ast }(\psi _{\mathrm{in},12})|0_{2}\rangle \otimes |0_{3}\rangle .  \label{july19_2c}
\end{equation}
That is, the second channel has one photon while both the first and third
channels are in the vacuum state.
\end{example}

\subsection{State transfer}\label{subsec:mN_output}

In this subsection, we derive an analytic form of the steady-state output field 
state of a passive quantum linear system driven by an $m$-channel $N$-photon input 
state defined in Eq. (\ref{eq:temp1}).

The following is the main result of this section.

\begin{theorem}\label{thm:general} 
Let the asymptotically stable passive quantum
linear system (\ref{sys_passive}) be initialized in the vacuum state and driven by the $m$-channel $N$-photon input $|\Psi _{\mathrm{in}}\rangle $ defined in Eq. (\ref{eq:temp1}). The steady-state output field state is another $m$-channel $N$-photon state of the form 
\begin{equation}
|\Psi _{\mathrm{out}}\rangle 
=
\tprod\limits_{j=1}^{N}\sum_{l=1}^{m}\int_{-\infty }^{\infty }dt\ \psi _{\mathrm{out},jl}(t)b_{\mathrm{in},l}^{\ast }(t)|0^{\otimes m}\rangle ,
\label{eq:temp5}
\end{equation}
where the output pulses are given by
\begin{equation}
\psi _{\mathrm{out},jl}(t)\triangleq \sum_{k=1}^{m}\int_{-\infty }^{\infty
}g_{G^{-}}^{lk}(t-r)\psi _{\mathrm{in},jk}(r)dr,\ \ \ j=1,\ldots ,N,  \  l=1,\ldots ,m.  \label{eq:temp4}
\end{equation}
\end{theorem}

\textbf{Proof.~} The proof is similar to that for Theorem \ref{thm:passive_mm}; specifically,
\begin{eqnarray}
&&
|\Psi _{\mathrm{out}}\rangle  
\nonumber
\\
&=&
\lim_{t_{0}\rightarrow -\infty ,t\rightarrow \infty }\left\langle \phi
\right\vert U(t,t_{0})|\phi \Psi _{\mathrm{in}}\rangle  
\nonumber
\\
&=&
\lim_{t_{0}\rightarrow -\infty ,t\rightarrow \infty }\left\langle \phi \right\vert
U(t,t_{0})\tprod\limits_{j=1}^{N}\sum_{k=1}^{m}\int_{-\infty }^{\infty }dt_k\
\psi _{\mathrm{in},jk}(t_{k})b_{\mathrm{in},k}^{\ast }(t_{k})|\phi
0^{\otimes m}\rangle  
\nonumber
\\
&=&
\lim_{t_{0}\rightarrow -\infty ,t\rightarrow \infty }\tprod\limits_{j=1}^{N}\sum_{k=1}^{m}\left\langle \phi
\right\vert \int_{t_{0}}^{t}dt_{k}\ \psi _{\mathrm{in},jk}(t_{k})U(t_{k},t_{0})b_{\mathrm{in},k}^{\ast
}(t_{k})U(t_{k},t_{0})^{\ast }|\phi 0^{\otimes m}\rangle  
\nonumber
\\
&=&
\lim_{t\rightarrow \infty }
\tprod\limits_{j=1}^{N}\sum_{k=1}^{m}\left\langle \phi \right\vert
\int_{-\infty }^{t}dt_{k}\ \psi _{\mathrm{in},jk}(t_{k})U(t_{k},-\infty )b_{\mathrm{in},k}^{\ast }(t_{k})U(t_{k},-\infty )^{\ast }|\phi 0^{\otimes
m}\rangle 
\nonumber
 \\
&=&
\tprod\limits_{j=1}^{N}\sum_{k=1}^{m}
\left\langle \phi \right\vert \int_{-\infty }^{\infty }dt_{k}\ \psi _{\mathrm{in},jk}(t_{k})\int_{-\infty }^{\infty
}\sum_{l=1}^{m}g_{G^{-}}^{lk}(r_{k}-t_{k})b_{\mathrm{in},l}^{\ast
}(r_{k})dr_{k}|\phi 0^{\otimes m}\rangle  
\nonumber
\\
&=&
\tprod\limits_{j=1}^{N}\sum_{l=1}^{m}
\int_{-\infty }^{\infty }dr_{k}\ \boxed{\sum_{k=1}^{m}\int_{-\infty }^{\infty
}dt_{k}\ g_{G^{-}}^{lk}(r_{k}-t_{k})\psi _{\mathrm{in},jk}(t_{k})} \ b_{\mathrm{in},l}^{\ast }(r_{k})|0^{\otimes m}\rangle  
\nonumber
\\
&=&
\tprod\limits_{j=1}^{N}\sum_{l=1}^{m}
\int_{-\infty }^{\infty }dr_k\ \boxed{\psi _{\mathrm{out},jl}(r_k)}\ b_{\mathrm{in},l}^{\ast }(r_k)|0^{\otimes m}\rangle ,
\label{eq:jan15_final}
\end{eqnarray}
where the output pulse shapes $\psi _{\mathrm{out},jl}$ are those 
in Eq. (\ref{eq:temp4}), as highlighted by the two boxes above. $\blacksquare $

\section{Conclusion}\label{sec:conclusion} 
In this paper, we have studied the dynamics of
quantum linear systems in response to multi-channel multi-photon states. We
have derived the intensity of the output field which can be used to
investigate the influence of quantum linear systems on quantum correlations
of multi-photon light fields. We have also presented the explicit formula of the
steady-state output field states when a quantum linear system is driven by three  classes of multi-channel
multi-photon input states. The results presented here are very general and
hold promising applications in photon-based quantum coherent feedback networks. One of the future research directions is to study controller synthesis problem on the basis of the system analysis carried out in this paper,  for example, via the Lyapunov method \cite{MvH07}, \cite{KC08}.

\begin{ack}                               
The author wishes to thank the anonymous reviewers for their careful reading and constructive comments.
\end{ack}

\bibliographystyle{plain}
\bibliography{gzhang}

\begin{thebibliography}{10}

\bibitem{AT12}
C.~Altafini and F.~Ticozzi.
\newblock Modeling and control of quantum systems: an introduction.
\newblock {\em IEEE Trans. Automat. Contr.}, 57:1898--1917, 2012.

\bibitem{AM79}
B.~D.~O. Anderson and J.~B. Moore.
\newblock {\em Optimal Filtering}.
\newblock Prentice-Hall, Englewood Cliffs, NJ, 1979.

\bibitem{BR04}
H.-A. Bachor and T.~C. Ralph.
\newblock {\em A Guide to Experiments in Quantum Optics}.
\newblock Wiley, 2004.

\bibitem{BCB+12}
B.~Q. Baragiola, R.~L. Cook, A.~M. Branczyk, and J.~Combes.
\newblock N-photon wave packets interacting with an arbitrary quantum system.
\newblock {\em Phys. Rev. A.}, 86:013811, 2012.

\bibitem{BDS+12}
T.~J. Bartley, G.~Donati, J.~B. Spring, X.~M. Jin, M.~Barbieri, A.~Datta, B.~J.
  Smith, and I.~A. Walmsley.
\newblock Multiphoton state engineering by heralded interference between single
  photons and coherent states.
\newblock {\em Phys. Rev. A}, 86:043820, 2012.

\bibitem{VPB80}
V.~P. Belavkin.
\newblock Quantum filtering of markov signals with white quantum noise.
\newblock {\em Radiotechnika i Electronika}, 25:1445--1453, 1980.

\bibitem{VPB83}
V.~P. Belavkin.
\newblock On the theory of control of observable quantum systems.
\newblock {\em Automat. Rem. Control}, 44:178--188, 1983.

\bibitem{VPB92}
V.~P. Belavkin.
\newblock Quantum stochastic calculus and quantum nonlinear filtering.
\newblock {\em J. Multivariate Anal.}, 42:171--201, 1992.

\bibitem{VPB93}
V.~P. Belavkin.
\newblock Quantum diffusion, measurement and filtering.
\newblock {\em Theory Probab. Appl.}, 38:573--585, 1993.

\bibitem{Bracewell99}
R.~N. Bracewell.
\newblock {\em The Fourier Transform and its Applications,3 edition}.
\newblock McGraw Hill, 1999.

\bibitem{BDS+15}
B.~Brecht, Dileep~V. Reddy, C.~Silberhorn, and M.~G. Raymer.
\newblock Photon temporal modes: A complete framework for quantum information
  science.
\newblock {\em Phys. Rev. X}, 5:041017, Oct 2015.

\bibitem{BH97}
R.~G. Brown and P.~Y.~C. Hwang.
\newblock {\em Introduction to Random Signals and Applied Kalman Filtering, 3rd
  ed,}.
\newblock John Wiley \& Sons, 1997.

\bibitem{CHJ12}
A.~R.~R. Carvalho, M.~R. Hush, and M.~R. James.
\newblock Cavity driven by a single photon: conditional dynamics and nonlinear
  phase shift.
\newblock {\em Phys. Rev. A.}, 86:023806, 2012.

\bibitem{CMR09}
J.~Cheung, A.~Migdall, and M.~L. Rastello.
\newblock Special issue on single photon sources, detectors, applications, and
  measurement methods.
\newblock {\em J. Modern Optics}, 56:139--140, 2009.

\bibitem{DP10}
D.~Dong and I.~R. Petersen.
\newblock Quantum control theory and applications:a survey.
\newblock {\em IET Control Theory \& Applications}, 4:2651--2671, 2010.

\bibitem{FKS10}
S.~Fan, S.~E. Kocabas, and J.-T. Shen.
\newblock Input-output formalism for few-photon transport in one-dimensional
  nanophotonic waveguides coupled to a qubit.
\newblock {\em Phys. Rev. A}, 82:063821, Dec 2010.

\bibitem{GZ00}
C.~W. Gardiner and P.~Zoller.
\newblock {\em Quantum Noise: A Handbook of Markovian and Non-Markovian Quantum
  Stochastic Methods with Applications to Quantum Optics}.
\newblock Springer, 2004.

\bibitem{GEP+98}
K.~M. Gheri, K.~Ellinger, T.~Pellizzari, and P.~Zoller.
\newblock Photon-wavepackets as flying quantum bits.
\newblock {\em Fortschr. Phys.}, 46:401--415, 1998.

\bibitem{GJ09}
J.~E. Gough and M.~R. James.
\newblock The series product and its application to quantum feedforward and
  feedback networks.
\newblock {\em IEEE Trans. Automat. Contr.}, 54:2530--2544, 2009.

\bibitem{GJN10}
J.~E. Gough, M.~R. James, and H.~I. Nurdin.
\newblock Squeezing components in linear quantum feedback networks.
\newblock {\em Phys. Rev. A}, 81:023804, 2010.

\bibitem{GJN11}
J.~E. Gough, M.~R. James, and H.~I. Nurdin.
\newblock Quantum filtering for systems driven by fields in single photon
  states and superposition of coherent states using non-markovian embeddings.
\newblock {\em Quantum Information Processing}, 12:1469--1499, 2013.

\bibitem{GJNC12}
J.~E. Gough, M.~R. James, H.~I. Nurdin, and J.~Combes.
\newblock Quantum filtering for systems driven by fields in single-photon
  states or superposition of coherent states.
\newblock {\em Phys. Rev. A}, 86:043819, Oct 2012.

\bibitem{GY16}
M.~Guta and N.~Yamamoto.
\newblock System identification for passive linear quantum systems.
\newblock {\em IEEE Transactions on Automatic Control}, 61(4):921--936, April
  2016.

\bibitem{HP84}
R.~L. Hudson and K.~R. Parthasarathy.
\newblock Quantum ito's formula and stochastic evolutions.
\newblock {\em Communications in Mathematical Physics}, 93(3):301--323, 1984.

\bibitem{JG10}
M.~R. James and J.~E. Gough.
\newblock Quantum dissipative systems and feedback control design by
  interconnection.
\newblock {\em IEEE Trans. Automat. Control}, 55:1806--1821, 2010.

\bibitem{KB09}
T.~G. Kolda and B.~W. Bader.
\newblock Tensor decompositions and applications.
\newblock {\em SIAM Review}, 51:455--500, 2009.

\bibitem{KC08}
S.~Kuang and S.~Cong.
\newblock Lyapunov control methods of closed quantum systems.
\newblock {\em Automatica}, 44(1):98 -- 108, 2008.

\bibitem{KS72}
H.~Kwakernaak and R.~Sivan.
\newblock {\em Linear Optimal Control Systems}.
\newblock John Wiley and Sons, Inc, 1972.

\bibitem{Leo03}
U.~Leonhardt.
\newblock Quantum physics of simple optical instruments.
\newblock {\em Rep. Prog. Phys.}, 66:1207--1249, 2003.

\bibitem{LL13}
L.-Q. Liao and C.~K. Law.
\newblock Correlated two-photon scattering in cavity optomechanics.
\newblock {\em Phys. Rev. A}, 87:043809, Apr 2013.

\bibitem{RL00}
R.~Loudon.
\newblock {\em The Quantum Theory of Light, 3rd~ed}.
\newblock Oxford University Press, Oxford, 2000.

\bibitem{Milburn08}
G.~J. Milburn.
\newblock Coherent control of single photon states.
\newblock {\em Eur. Phys. J. Special Topics}, 159:113--117, 2008.

\bibitem{MvH07}
M.~Mirrahimi and R.~van Handel.
\newblock Stabilizing feedback controls for quantum systems.
\newblock {\em SIAM J. Control and Optim.}, 46:445--467, 2007.

\bibitem{MNM10}
W.~J. Munro, K.~Nemoto, and G.~J. Milburn.
\newblock Intracavity weak nonlinear phase shifts with single photon driving.
\newblock {\em Optics Communications}, 283:741--746, 2010.

\bibitem{NC00}
M.~A. Nielsen and I.~L. Chuang.
\newblock {\em Quantum Computation and Information}.
\newblock Cambridge University Press, London, 2000.

\bibitem{HIN14}
H.~I. Nurdin.
\newblock Structures and transformations for model reduction of linear quantum
  stochastic systems.
\newblock {\em Automatic Control, IEEE Transactions on}, 59(9):2413--2425, Sept
  2014.

\bibitem{NJD09}
H.~I. Nurdin, M.~R. James, and A.~C. Doherty.
\newblock Network synthesis of linear dynamical quantum stochastic systems.
\newblock {\em SIAM J. Control and Optim}, 48:2686--2718, 2009.

\bibitem{NKM+15}
A.~Nysteen, P.~T. Kristensen, D.~P.~S. McCutcheon, P.~Kaer, and J.~Mazrk.
\newblock Scattering of two photons on a quantum emitter in a one-dimensional
  waveguide: exact dynamics and induced correlations.
\newblock {\em New Journal of Physics}, 17(2):023030, 2015.

\bibitem{Ou07}
Z.~Y. Ou.
\newblock Multi-photon interference and temporal distinguishability of photons.
\newblock {\em Int. J. Modern Physics B}, 21:5033--5058, 2007.

\bibitem{PZJ15}
Y.~Pan, G.~Zhang, and M.~R. James.
\newblock Analysis and control of quantum finite-level systems driven by
  single-photon input states.
\newblock {\em Automatica}, 69:18--23, 2016.

\bibitem{IRP11}
I.~R. Petersen.
\newblock Cascade cavity realization for a class of complex transfer functions
  arising in coherent quantum feedback control.
\newblock {\em Automatica}, 47(8):1757 -- 1763, 2011.

\bibitem{QSW07}
L.~Qi, W.~Sun, and Y.~Wang.
\newblock Numerical multilinear algebra and its applications.
\newblock {\em Front. Math. China}, 2:501--526, 2007.

\bibitem{QZ09}
L.~Qiu and K.~Zhou.
\newblock {\em Introduction to Feedback Control}.
\newblock Prentice-Hall, 2009.

\bibitem{RMS07}
P.~P. Rohde, W.~Mauerer, and C.~Silberhorn.
\newblock Spectral structure and decompositions of optical states, and their
  applications.
\newblock {\em New Journal of Physics}, 9(4):91, 2007.

\bibitem{Sog10}
T.~Sogo.
\newblock On the equivalence between stable inversion for nonminimum phase
  systems and reciprocal transfer functions defined by the two-sided laplace
  transform.
\newblock {\em Automatica}, 46:122--126, 2010.

\bibitem{SZX15}
H.~Song, G.~Zhang, and Z.~R. Xi.
\newblock Continuous-mode multi-photon filtering.
\newblock {\em SIAM Journal on Control and Optimization}, 54:1602--1632, 2016.

\bibitem{vHSM05}
R.~van Handel, J.~K. Stockton, and M.~Mabuchi.
\newblock Feedback control of quantum state reduction.
\newblock {\em IEEE Trans. Automat. Contr.}, 50:768--780, 2005.

\bibitem{WM08}
D.~F. Walls and G.~J. Milburn.
\newblock {\em Quantum Optics, 2nd~ed}.
\newblock Springer, 2008.

\bibitem{WM10}
H.~W. Wiseman and G.~J. Milburn.
\newblock {\em Quantum Measurement and Control}.
\newblock Cambridge University Press, Cambridge, UK, 2010.

\bibitem{NY13}
N.~Yamamoto.
\newblock Decoherence-free linear quantum subsystems.
\newblock {\em IEEE Transactions on Automatic Control}, 59(7):1845--1857, 2014.

\bibitem{YJ14}
N.~Yamamoto and M.~R. James.
\newblock Zero-dynamics principle for perfect quantum memory in linear
  networks.
\newblock {\em New Journal of Physics}, 16(7):073032, 2014.

\bibitem{YK03a}
M.~Yanagisawa and H.~Kimura.
\newblock Transfer function approach to quantum control-part i: dynamics of
  quantum feedback systems.
\newblock {\em IEEE Trans. Automat. Contr.}, 48:2107--2120, 2003.

\bibitem{YK03b}
M.~Yanagisawa and H.~Kimura.
\newblock Transfer function approach to quantum control-part ii: Control
  concepts and applications.
\newblock {\em IEEE Transactions on Automatic Control}, 48(12):2121--2132, Dec
  2003.

\bibitem{Yukawa:13}
M.~Yukawa, K.~Miyata, T.~Mizuta, H.~Yonezawa, P.~Marek, R.~Filip, and
  A.~Furusawa.
\newblock Generating superposition of up-to three photons for continuous
  variable quantum information processing.
\newblock {\em Opt. Express}, 21(5):5529--5535, Mar 2013.

\bibitem{zhang14}
G.~Zhang.
\newblock Analysis of quantum linear systems' response to multi-photon states.
\newblock {\em Automatica}, 50:442--451, 2014.

\bibitem{zhang16}
G.~Zhang, S.~Grivopoulos, I.~R. Petersen, and J.~E. Gough.
\newblock On the structure of quantum linear systems.
\newblock {\em arXiv:1606.05719}, 2016.

\bibitem{zhang11}
G.~Zhang and M.~R. James.
\newblock Direct and indirect couplings in coherent feedback control of linear
  quantum systems.
\newblock {\em IEEE Trans. Automat. Contr.}, 56:1535--1550, 2011.

\bibitem{zhang12}
G.~Zhang and M.~R. James.
\newblock Quantum feedback networks and control: a brief survey.
\newblock {\em Chinese Science Bulletin}, 57:2200--2214, 2012.

\bibitem{zhang13}
G.~Zhang and M.~R. James.
\newblock On the response of quantum linear systems to single photon input
  fields.
\newblock {\em IEEE Trans. Automat. Contr.}, 58:1221--1235, 2013.

\bibitem{ZLW+15}
J.~Zhang, Y-X Liu, R-B Wu, K.~Jacobs, and F.~Nori.
\newblock Quantum feedback: theory, experiments, and applications.
\newblock {\em arXiv:1407.8536v3 [quant-ph]}, 2015.

\bibitem{ZWL12}
J.~Zhang, R-B Wu, Y-X Liu, C-W Li, and T-J Tarn.
\newblock Quantum coherent nonlinear feedbacks with applications to quantum
  optics on chip.
\newblock {\em IEEE Trans. Automat. Contr.}, 57:1997--2008, 2012.

\bibitem{ZDG96}
K.~Zhou, J.~C. Doyle, and K.~Glover.
\newblock {\em Robust and Optimal Control}.
\newblock Prentice-Hall, Upper Saddle River, NJ, 1996.

\end{thebibliography}

\end{document}